\documentclass[nohyper]{JHEP3}
\usepackage{amsmath,amssymb,amsfonts,epsfig,undertilde}

\title{5D Black Holes and Non-linear Sigma Models}

\author{Micha Berkooz\\
Department of Particle Physics,
The Weizmann Institute of Science, Rehovot 76100,
Israel\\
\\
{\tt E-mail:berkooz@wisemail.weizmann.ac.il}}

\author{Boris Pioline\\
Laboratoire de Physique Th\'eorique et Hautes
Energies\footnote{Unit\'e mixte de recherche du CNRS UMR 7589},\\
Universit\'e Pierre et Marie Curie - Paris 6,
4 place Jussieu, F-75252 Paris cedex 05 \\

Laboratoire de Physique Th\'eorique de l'Ecole Normale
Sup\'erieure\footnote{Unit\'e mixte
de recherche du CNRS UMR 8549},\\
24 rue Lhomond, F-75231 Paris cedex 05\\
\\
{\tt E-mail: pioline@lpthe.jussieu.fr}
}

\abstract{Stationary solutions of 5D supergravity with $U(1)$ isometry
can be efficiently studied by dimensional reduction to three dimensions,
where they reduce to solutions to a locally supersymmetric non-linear
sigma model. We generalize this procedure to 5D gauged supergravity,
and identify the corresponding gauging in 3D. We pay particular
attention to the case where the Killing spinor is non constant along
the fibration, which results, even for ungauged supergravity in 5D,
in an additional gauging in 3D, without introducing any extra potential.
We further study $SU(2)\times U(1)$ symmetric solutions, which correspond 
to geodesic motion on the sigma model (with potential in the gauged case). 
We identify and study the algebra of BPS constraints relevant for the 
Breckenridge-Myers-Peet-Vafa black hole, the Gutowski-Reall 
black hole and several other BPS solutions, and obtain the corresponding 
radial wave functions in the semi-classical approximation.
}

\preprint{LPTENS-08-11\\WIS/05/08-FEB-DPP}

\begin{document}

\renewcommand{\Im}{{\rm Im}}
\renewcommand{\Re}{{\rm Re}}
\newcommand{\pa}{\partial}
\newcommand{\nn}{\nonumber}
\newcommand{\eps}{\epsilon}
\newcommand{\IR}{\mathbb{R}}
\newcommand{\IC}{\mathbb{C}}
\newcommand{\IZ}{\mathbb{Z}}
\newcommand{\IH}{\mathbb{H}}
\newcommand{\IO}{\mathbb{O}}
\newcommand{\cN}{\mathcal{N}}
\newcommand{\cM}{\mathcal{M}}
\newcommand{\cX}{\mathcal{X}}
\newcommand{\tzeta}{{\tilde\zeta}}
\newcommand{\I}{{\rm i}}
\newcommand{\J}{{\rm j}}
\def\bea{\begin{eqnarray}}
\def\eea{\end{eqnarray}}
\def\be{\begin{equation}}
\def\ee{\end{equation}}
\def\ba{\begin{align}}
\def\ea{\end{align}}
\def\bse{\begin{subequations}}
\def\ese{\end{subequations}}


\section{Introduction}

While mostly of theoretical interest, BPS solutions of 5D gravity
and supergravity have been the subject of intense studies recently,
partly due to their unexpected variety (see e.g. \cite{Emparan:2008eg}
for a recent review), to the relative simplicity
of their microscopic dynamics~\cite{Strominger:1996sh,
Breckenridge:1996is}, to their relation to 4D BPS black holes via
the 4D/5D lift \cite{Gaiotto:2005gf,Gaiotto:2005xt}, and, in the
case of gauged supergravity, to their relevance for the dynamics
of four-dimensional gauge theories via the AdS/CFT correspondance.
The purpose of this work is to develop algebraic techniques
for constructing 5D BPS black hole solutions, generalizing existing methods
which have been successfully applied to 4D black holes.

An important motivation for our study is the supersymmetric $AdS_5$ black
hole solution found by Gutowski and Reall (GR) \cite{Gutowski:2004ez,
Gutowski:2004yv}, which has remained particularly mysterious: for example,
the solution exhibits a certain relation between the angular
momentum and the electric charges, which is not implied by the $N=4$
superconformal algebra on the boundary. This restriction remains
in generalizations involving two different angular
momenta \cite{Chong:2005da,Kunduri:2006ek}. This situation is in contrast
with the asymptotically flat 5D space-time, where BPS solutions exist for
arbitrary values of the angular momenta and charges within certain bounds.
It is an open problem whether more general $AdS_5$ solutions exist
where the above restriction is relaxed (see \cite{Kunduri:2006uh,
Kunduri:2007qy} for recent progress on this issue).

Moreover, no microscopic
counting of the entropy of the Gutowski-Reall black hole
from the dual $N=4$ SYM theory is available
to date. While 1/8-BPS (or more) supersymmetric states can be counted on
the gauge theory side at weak coupling using a suitable
index \cite{Kinney:2005ej}, 1/16-BPS states in general pair up due to the
interactions, and the resulting (order $N$) index is much smaller than
the (order $N^2$) entropy of a large GR black hole.
Understanding this problem in more detail would be a useful
step in bridging the gap between  our remarkable control over black holes
with high SUSY, and our qualitative understanding, at best, of
non-supersymmetric black holes.

A heuristic model identifying
a class of fermionic operators in the gauge theory which reproduces
some of the scaling properties of the GR ($AdS_5$) black hole was put 
forward in \cite{Berkooz:2006wc}.
For 1/16-BPS black holes in $AdS_4$ \cite{Duff:1999gh,Cvetic:2005zi}, 
the same model suggests that
the entropy counts $N^{3/2}$ degrees of freedom.
This may provide a supersymmetric setting to study the long-standing problem
of counting the entropy of 2+1 dimensional strongly coupled fixed points.

In the flat four-dimensional case, the integrable structure
of stationary solutions exposed by the dimensional reduction to
three dimensions proved very useful in mapping the phase diagram
of black holes. In this paper we generalize this
algebraic description to stationary solutions of
five-dimensional $\cN=1$ supergravity with an additional $U(1)$ symmetry.
Thus, we assume two commuting Killing vectors $\pa_t$ and $\pa_{\psi}$,
time-like and space-like, respectively.
The description in three dimensions is given in terms
of a non-linear sigma model on (an analytic continuation of)
a quaternionic-K\"ahler manifold $\mathcal{M}_3$,
coupled to Euclidean gravity in three dimensions.

For explicitness, we focus for the most part
on minimal supergravity in 5D, which
leads to a non-linear sigma model on $G_3/K_3=G_{2(2)}/SO(4)$
(see \cite{Bodner:1989cg,Mizoguchi:1998wv} for early discussions of 
this model, and \cite{Bouchareb:2007ax,Clement:2007qy} for an independent
study of its application to 5D black holes in ungauged supergravity).
The same sigma model (up to analytic continuation) has appeared in the
study of 4D black holes \cite{Gaiotto:2007ag}, and is in fact
related to the present one by a flip of the $t$ and $\psi$ directions,
corresponding to a Weyl reflection in a $Sl(2)$ subgroup of $G_3$.
The quaternionic geometry of $G_{2(2)}/SO(4)$ was studied in great detail
in \cite{Gunaydin:2007qq}, whose notations we follow.

A crucial difference with
the dimensional reduction of 4D BPS black holes \cite{Gunaydin:2005mx,
Neitzke:2007ke,Gunaydin:2007bg,Pioline:2006ni} is the fact that
covariantly constant spinors in 5D are in general not constant along the
orbit of the space-like Killing vector $\pa_\psi$.
As a result, the 3D sigma model including the fermions
is gauged, even when the 5D supergravity is ungauged.
This gauging does not
affect the bosonic part of the action, however.
When the 5D supergravity is gauged,
an additional potential is generated in the 3D sigma model.
We identify the correct gauging in 
the general framework for 3D gauged supergravities laid out in 
\cite{de Wit:2003ja,deWit:2004yr}.

When a further $SU(2)$
symmetry is present, the model may be further reduced to one
dimension, where it reduces to geodesic motion of a fictitious
particle on a real cone over $\mathcal{M}_3$.
The gauging in 5D introduces a potential,
which spoils the integrability of the model in general.
Some of the algebraic structure however does carry over,
and determines the structure of the BPS constraints.

As a by-product of this analysis, we obtain the semi-classical
form of the radial wave function for 5D BPS black holes, i.e.
the solution to the BPS constraints in Hamilton-Jacobi formalism.

The outline of the paper is as follows. In Section 2 we discuss
the dimensional reduction of 5D $\cN=1$ supergravity down to 4,
3 and 1 dimensions. We identify the gauging of the 3D sigma model
coming from the dependence of the Killing spinor along the direction
$\psi$, and from the gauging in 5 dimensions. In Section 3
we specialize to stationary solutions with $U(1)\times SU(2)$
isometries in minimal supergravity in 5D. We identify the
algebraic structure of the supersymmetry constraints 
appropriate to the BMPV, Taub-NUT,
G\"odel, Eguchi-Hanson and Gutowski-Reall black holes, respectively,
and compute their respective Noether charges and radial wave functions.
Section 4 contains a brief summary and discussion. A particular class
of solutions with nilpotent Noether charges of degree 2 is presented
in Appendix A.

\section{5D Black holes and non-linear sigma models}
We consider $\cN=1$, $D=4+1$ supergravity coupled to $n$ abelian
vector multiplets. The bosonic part of the Lagrangian is given by
\be
\label{lag5d}
e^{-1}\mathcal{L}_5=-\frac12 R - \frac14 G_{IJ} F_{\mu\nu}^I F^{\mu\nu J}
-\frac12 g_{ij} \pa_\mu \varphi^i \pa^\mu \varphi^j + \frac{e^{-1}}{48}
\epsilon^{\mu\nu\rho\sigma\lambda} C_{IJK} F_{\mu\nu}^I F_{\rho\sigma}^J
A_\lambda^K
\ee
The scalars $\varphi^i$ take value in the moduli space $\mathcal{M}_5$,
given by the cubic hypersurface
\be
\label{vone}
I_3(h) \equiv \frac16 C_{IJK} h^I h^J h^K = 1
\ee
where $C_{IJK}$ are constants.
The metric for the kinetic terms of the scalars $\varphi^i$ and
the gauge fields $A^I$ are given by
\be
g_{ij}=G_{IJ}\pa_{\varphi^i} h^I \pa_{\varphi^j} h^J\ ,\quad
G_{IJ} = -\frac12 \pa_{h^I} \pa_{h^J} \log I_3(h)\ ,\quad
\ee
evaluated on the hypersurface \eqref{vone}.
For simplicity, we restrict to the case where the moduli space
is a symmetric space, so that~\cite{Gunaydin:1983bi}
\be
C_{IJK} C_{J'(LM} C_{PQ)K'} \delta^{JJ'}
\delta^{KK'}=\frac43 \delta_{I(L}C_{MNP)}
\ee
In this case $I_3(h)$ is the norm form of a Jordan algebra $J$ of
degree 3 \cite{Gunaydin:1984ak}, $\mathcal{M}_5=G_5/K_5=
{\rm Str}_0(J)/{\rm Aut}(J)$
where ${\rm Str}_0(J)$ and ${\rm Aut}(J)$ are the reduced structure group
and automorphism groups of $J$, and
\be
C^{IJK}\equiv \delta^{II'} \delta^{JJ'} \delta^{KK'} C_{I'J'K'} \ .
\ee
It is useful to define the ``adjoint map''
\be
h_I^\sharp \equiv \frac16 C_{IJK} h^J h^K\ ,\quad
\ee
which satisfies
\be
h^I = \frac92 C^{IJK} h^\sharp_J h^\sharp_K
\ee
In cases where the moduli space is not symmetric, the reduction procedure
that we shall describe below still applies, however the resulting
moduli space in three dimensions is no longer symmetric.

In the absence of
hypermultiplets, it is possible to include a Fayet-Iliopoulos term
for a linear combination ${\cal A}_\mu=V_I A^I_\mu$ of the $n$ gauge
fields ($V_I$ are numerical constants). The Lagrangian becomes
$e^{-1}{\cal L}_{5;{\rm gauged}}=e^{-1}{\cal L}_5+V_5$, where
the potential is given by~\cite{Gunaydin:1984ak,Gunaydin:1986fh}
\be
\label{V5}
V_5=g^2 V_I V_J \left( 6 h^I h^J - \frac92 g^{ij} \pa_i h^I \pa_j h^J \right)
=27 C^{IJK} V_I V_J h_K
\ee
The potential admits an $AdS_5$ vacuum provided $V_I$ lies inside the
cone $I_3(V)>0$.

\subsection{Stationary solutions}
Assuming the existence of a time-like Killing vector, the five-dimensional
metric and gauge fields can be taken in the form
\be
\label{an410}
ds^2_5 = -f^2 (dt+\omega_4)^2 + f^{-1} ds^2_4 \ ,\quad
A_5^I = \phi^I (dt+\omega_4) + A_4^I
\ee
where $f$, $\phi^I$ and $\varphi^i$ are independent
of time, and $A_4^I, \omega_4$ are one-forms on the four-dimensional
Euclidean slice. The equations of motion for this ansatz are most easily
obtained by reducing the Lagrangian along the time direction.
This leads to $\cN=2$ supergravity in $D=4$ Euclidean dimensions,
coupled to $n+1$ vector multiplets. The reduced Lagrangian $\mathcal{L}_4$
is determined in the usual way by the holomorphic prepotential
\be
F = \frac16 C_{IJK} X^I X^J X^K / X^0
\ee
In constrast to the usual Kaluza-Klein reduction along a space-like
direction, studied for example in \cite{Gunaydin:2005df},
the special coordinates $z^I=X^I/X^0$ and
$\bar z^I=\bar X^I/\bar X^0$ are independent real variables,
\be
z^I = \frac{X^I}{X^0} = \phi^I + {\cal I}\,f\,h^I\ ,\quad
\bar z^I = \phi^I - {\cal I}\, f\, h^I
\ee
where ${\cal I}^2=-1,\bar {\cal I}=-{\cal I}$
is a ``para-complex'' structure ~\cite{Cortes:2003zd}.
As a result, the vector moduli space $\mathcal{M}_4^*$ has split
signature $(n+1,n+1)$.
In the following, we will perform an analytic continuation
$\phi^I\to i \phi^I$, which allows us to work with the
standard complex structure ${\cal I}=i$, 
albeit with a purely imaginary $\phi^I$.
Similarly, we shall continue  $\omega_4 \to i\omega_4$, so that
$A^\Lambda_4 = (A^0_4, A^I_4) = (\omega_4, A_4^I)$
and their magnetic duals transform as a vector of $Sp(2n+2)$.
For later reference, we note that the K\"ahler potential is given by
\be
K = -\log I_3(z^I-\bar z^I) = - 3\, \log f
\ee
When $I_3$ is the norm form of a Jordan algebra $J$, the vector
multiplet moduli space  is a symmetric space $\mathcal{M}_4
=G_4/K_4={\rm Conf}(J)/[U(1)
\times \widehat{\rm Str}_0(J)]$, where ${\rm Conf}(J)$ is the conformal
group of $J$ and $\widehat{\rm Str}_0(J)$ is the compact form of
the reduced structure group of $J$.

In the presence of Fayet-Iliopoulos terms in 5 dimensions, the
scalar potential \eqref{V5} leads to a potential $V_4$
in four dimensions,
\be
\label{V4}
V_4 = f^{-1} V_5
\ee
Using for example the identities found in \cite{Ceresole:2007rq},
one may check that \eqref{V4} is consistent with the general form
of the scalar potential induced by Fayet-Iliopoulos terms in
four dimensions~\cite{Cremmer:1984hj,Andrianopoli:1996vr},
\be
V_4 = g^2 \,\left( g^{i\bar j} f_i^\Lambda \, f_{\bar j}^{\Sigma}
-3 e^{K} \bar X^\Lambda X^\Sigma \right)\, \vec P_\Lambda \cdot \vec P_\Sigma
\ee
where $\vec P_\Lambda$ are the triplets of Fayet-Iliopoulos terms,
chosen as
\be
\vec P_0= 0 \ ,\quad \vec P_I = V_I\, \vec n\ ,\quad \vec n\cdot \vec n=1\ .
\ee
In the language of $\cN=1$ supergravity, this corresponds to a
superpotential $W=g\,V_I X^I$.

\subsection{Reduction to $\IR\times U(1)$ symmetric solutions}
We now restrict to solutions with a extra $U(1)$ isometry, generated by
a Killing vector $\pa_\psi$ on the four-dimensional spatial slice.
Accordingly, the spatial metric $ds_4^2$ decomposes as
\be
\label{ds4an}
ds_4^2 = e^{2U} (d\psi + \omega_3)^2 + e^{-2U} ds_3^2
\ee
while the gauge fields decompose as
\be
A_4^\Lambda= \zeta^\Lambda (d\psi + \omega_3) + A_3^\Lambda
\ee
The equations of motion for this ansatz can be obtained by
further reducing the four-dimensional Euclidean supergravity
along the space-like direction $\pa_\psi$. Upon dualizing
the gauge fields $A_3^\Lambda$ and $\omega_3$ into pseudo-scalars
$\tzeta_\Lambda$ and $\sigma$, one obtains $\cN=4$ supergravity
in three Euclidean dimensions coupled to a non-linear sigma
model, with Lagrangian\cite{de Wit:1992up}\footnote{The full
supersymmetry in the gauged case
can only be displayed by adding two auxiliary gauge fields with Chern-Simons
couplings, see Section \ref{sym3d} below.}
\be
e^{-1}\,\mathcal{L}_3 = -\frac12 R -\frac12 G_{ab}
\pa\varphi^a \pa\varphi^b  + V_3
\ee
The scalar potential $V_3$, present only in the case of gauged supergravity,
is given by the reduction of \eqref{V4},
\be
\label{V3}
V_3 = f^{-1} \,e^{-2U}\, V_5
\ee
The target space of the sigma model, which we shall denote by
$\cM_3$, is coordinatized by\footnote{The symbol
$\varphi$, used in \eqref{lag5d} to denote the scalars
in 5 dimensions, hereforth denotes the scalars in 3 dimensions.}
$\varphi^a=\{U,z^I,\bar z^I,\zeta^\Lambda,
\tzeta_\Lambda,\sigma\}$.
It is related to the more familiar
quaternionic-K\"ahler manifold $\mathcal{M}_{3,E}$
(known as the $c$-map of $\mathcal{M}_4$) arising
in the usual Kaluza-Klein reduction along space-like directions,
and with positive-definite metric \cite{Ferrara:1989ik}
\bea
\label{cmap}
ds_{\cM_{3,E}}^2 &=& dU^2 + g_{I\bar J} dz^I dz^{\bar J}
+ e^{-4U} \left( d\sigma - \tzeta_\Lambda d\zeta^\Lambda
+ \zeta^\Lambda d\tzeta_\Lambda\right)^2\\
&&\hspace*{-1cm}-\frac12 e^{-2U} \left[
(\Im\cN)_{\Lambda\Sigma}d\zeta^\Lambda d\zeta^\Sigma
+(\Im\cN)^{\Lambda\Sigma}\left(d\tzeta_\Lambda+(\Re\cN)_{\Lambda R}d\zeta^R\right)
\left(d\tzeta_\Sigma+(\Re\cN)_{\Sigma T}d\zeta^T\right) \right]\ ,\nn
\eea
by analytically continuing
\be
\label{rightcont}
\left( \phi^I, \zeta^0 , \zeta^I , \tzeta_I, \tzeta_0 , \sigma\right) \to
\left( i \phi^I, i \zeta^0 , \zeta^I , i \tzeta_I, \tzeta_0 , i \sigma\right)
\ .
\ee
For convenience, we shall be using the Riemannian metric \eqref{cmap},
but allow $\phi^I,\zeta^0,\tzeta_I,\sigma$ to be purely imaginary.
We note that an equally valid procedure would have been to perform the
reduction along the space-like Killing vector $\pa_\psi$ first, and
then along the time-like Killing vector $\pa_t$. This of course leads to
the same non-linear sigma model on $\cM_3$ in three Euclidean dimensions,
however the analytic continuation that relates it to the Riemanniann
manifold $\mathcal{M}_{3,E}$, in the variables appropriate to this
reduction, is now
\be
\label{othercont}
\left( \phi^I, \zeta^0 , \zeta^I , \tzeta_I, \tzeta_0 , \sigma\right) \to
\left( \phi^I, i \zeta^0 , i \zeta^I , i \tzeta_I, i \tzeta_0 , \sigma\right)
\ .
\ee
As we discuss  later in this section, the two analytic continuations
\eqref{rightcont} and \eqref{othercont} are in fact related by a Weyl
reflection in an
$Sl(2)$ subgroup of their isometry group, corresponding to the exchange of
the $t$ and $\psi$ direction.
It is also worthwhile to note that the same sigma-model arises when describing
stationary solutions in $D=3+1$ $\cN=2$ supergravity
\cite{Breitenlohner:1987dg,Gunaydin:2005mx,Pioline:2006ni,Neitzke:2007ke,
Gunaydin:2007bg}; as
we shall see however, the supersymmetry conditions corresponding to 5D black
holes differ from those pertaining to 4D black holes.

In general, the space $\cM_3$ admits a solvable algebra of isometries,
originating from the diffeomorphism and gauge symmetries in 5 dimensions:
the Killing vectors
\bse
\label{isoqk}
\be
p^\Lambda = \pa_{\tzeta_\Lambda} + \zeta^\Lambda \pa_\sigma\ ,\quad
q_\Lambda = \pa_{\zeta^\Lambda}-\tzeta_\Lambda \pa_\sigma\ ,\quad
k = \pa_\sigma
\ee
generate a Heisenberg algebra $[p^\Lambda,q_\Sigma]=
-2\delta^\Lambda_\Sigma\,k$; the generators
\be
T_I = \pa_{\phi^I} + \zeta^0 \pa_{\zeta^I}
- C_{IJK} \zeta^J \pa_{\tzeta_K} - \tzeta_I \pa_{\tzeta_0}
\ee
are nilpotent of degree 4, act symplectically on $(p^\Lambda,q_\Lambda)$,
and commute with $k$;
the non-compact generators
\be
H=-\pa_U-\zeta^\Lambda \pa_{\zeta^\Lambda}
-\tzeta_\Lambda \pa_{\tzeta_\Lambda} -2 \sigma\pa_{\sigma}\ ,
\ee
\be
D=-\frac12\left(
- 3 \zeta^0 \pa_{\zeta^0}
- \zeta^I \pa_{\zeta^I}
+ \tzeta_I \pa_{\tzeta_I}
+ 2\phi^I \pa_{\phi^I} + 2 t^I \pa_{t^I}
+ 3 \tzeta_0 \pa_{\tzeta_0} \right)
\ee
\ese
give a bi-grading of the nilpotent part of the algebra.
The presence of the potential
$V_3$ breaks the $(H,D)$ symmetry to $H-2D$, but leaves all
other generators above unbroken.

\EPSFIGURE{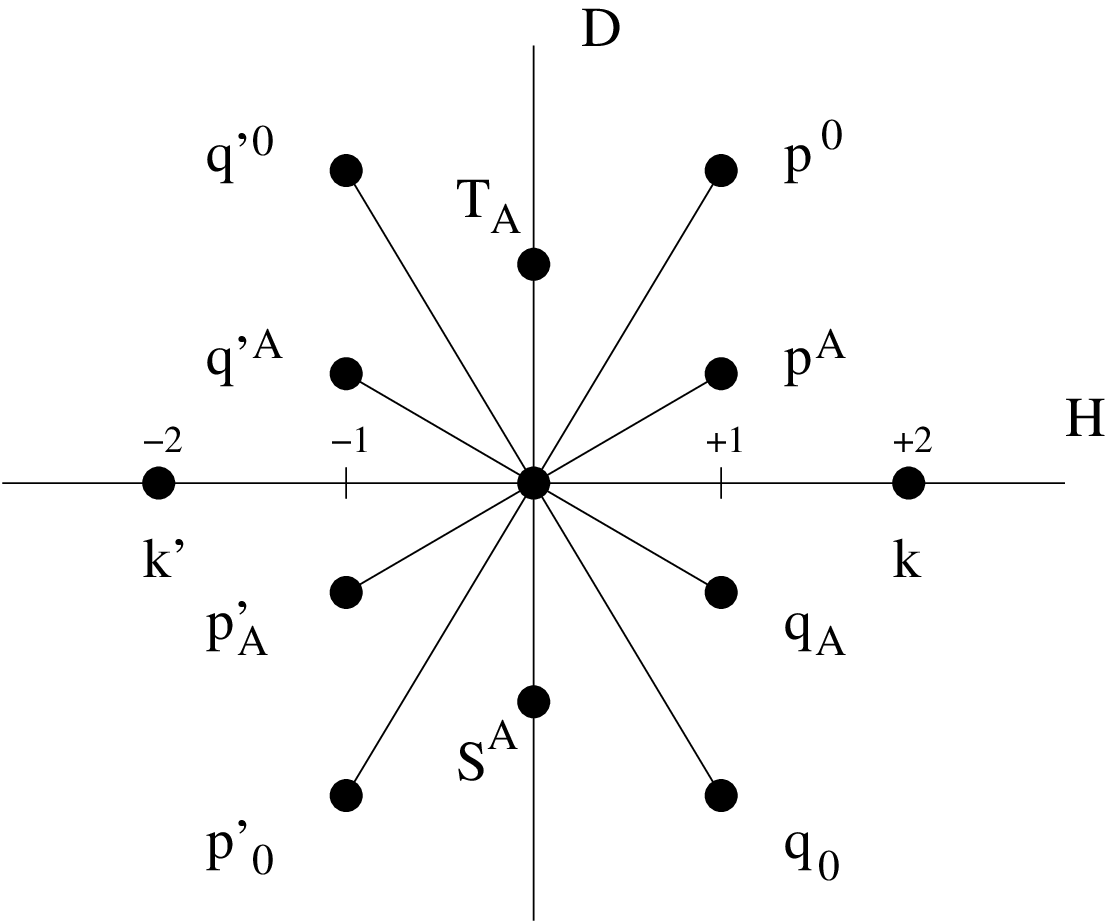, height=8cm}{Two-dimensional projection of the
root diagram of $G_3={\rm QConf}(J)$ with respect to the split Cartan
torus $(H,D)$. The long roots have multiplicity 1,
while the short roots have multiplicity $n+1$.\label{rootdiag}}

When $I_3$ is the norm form of a Jordan algebra $J$, the solvable
group of isometries is extended to a semi-simple group ${\rm QConf}(J)$,
such that $\mathcal{M}_3=G_3/K_3={\rm QConf}(J)/SU(2)_L\times 
\widetilde{\rm Conf}(J)$
becomes a quaternionic-K\"ahler symmetric space.
Here ${\rm QConf}(J)$ is the ``quasi-conformal group'' associated to
$J$ \cite{Gunaydin:2000xr,Gunaydin:2007qq} (see e.g. \cite{Pioline:2006ni}
for a review). It is obtained by 
supplementing the above generators with special transformations
$S^J$ and rotations $R^I_J$, such that $\{T_I,S^J,R_I^J,D\}$
generate $G_4={\rm Conf}(J)$, and with a ``dual'' Heisenberg algebra,
$[p_{'\Lambda},q^{'\Sigma}]=-2\delta_\Lambda^\Sigma\,k'$, requiring
that $\{k',H,k\}$ generate $Sl(2,\IR)$. The $SU(2)_L$ factor in
the maximal compact subgroup $K_3$ is the first factor in the 
R-symmetry group $SO(4)=
SU(2)_L \times SU(2)_R$, the scalars being inert under
the second factor $SU(2)_R$, which would act on the hypermultiplets
if those were present~\cite{de Wit:1992up}.

For later purposes, it will
be useful to recall that 
the root diagram of $G_3={\rm QConf}(J)$ admits a two-dimensional
projection given by the root diagram of the exceptional group $G_2$, where
the long roots have multiplicity one and the short roots have multiplicity
$n+1$ (see Figure \ref{rootdiag}). In particular, for minimal
supergravity in 5 dimensions, with $I_3(h)=h^3$ and $n=0$, the group $G_3$
is in fact $G_{2(2)}$ itself.
The long roots in Figure~\ref{rootdiag} 
generate a $Sl(3,\IR)$ subgroup of $G_3$,
which is the symmetry arising in the dimensional reduction of pure
Einstein gravity in $5$ dimensions down to 3
dimensions~\cite{Maison:1979kx,Giusto:2007fx,Ford:2007th}. In particular,
the $Sl(2,\IR)$ subgroup generated by the roots $q_0, q^{'0}$ and their
commutator is the symmetry exchanging the time-like and space-like
Killing vectors $\pa_t$ and $\pa_\psi$, alluded to
below \eqref{othercont}. The $Sl(2,\IR)$ subgroup generated by $k,k'$
and their commutator instead corresponds to the Ehlers symmetry of
Einstein gravity in 4 dimensions. In the presence of the potential
$V_3$, the only unbroken symmetries are $p^\Lambda, q_\Lambda, k$
and $T_I, q'_0$. As we shall see shortly, the conserved charges
associated to the latter are the electric charges and angular momentum
of the 5D black hole.

\subsection{Supersymmetry in 3 dimensions \label{sym3d}}

In the absence of gauging, the supersymmetry of the $N=4$ sigma
model coupled to gravity in three dimensions was discussed in
\cite{de Wit:1992up}. When gravity is gauged in 5 dimensions or when
the spinors are non-trivial along the fibers, then we obtain a
gauged model in three dimensions. In this subsection we will review
some aspects of gauged sigma models in three dimensions, which we will
use in subsequent sections. We will mainly follow the discussion in
\cite{de Wit:2003ja} and \cite{deWit:2004yr}.

\subsubsection*{The ungauged case}
A locally supersymmetric $\cN=4$ sigma model in 3 dimensions  
has an $SO(4)\sim SU(2)_L\times SU(2)_R$ R-symmetry. 
Out of this symmetry group, $SU(2)_R$
is already apparent in 5 dimensions where it acts on
hypermultiplets, leaving the bosonic fields in the vector multiplets
inert. The other factor $SU(2)_L$ is manifest only when reducing to 3D and
is associated to rotations in the two-plane of the fiber.
We use the following notations: a vector of the $SO(4)$ R-symmetry
carries an index $I=1\dots 4$, which is equivalent to a bi-spinor $\alpha
\dot \alpha$ ($\alpha=1,2, \dot\alpha=1,2$), where the dotted index
is the index for $SU(2)_L$. Indices of the adjoint of $SO(4)\equiv
SU(2)_L\times SU(2)_R$ will occasionally be denoted by $x=1,2,3$ and
${\dot x}=1,2,3$. In addition, $a,b,\dots$ will denote indices of
coordinates on the manifold. The $SO(4)$ R-symmetry determines an
$SO(4)$ connection, denoted by $Q^{[IJ]}_a$, or $Q^{\dot x}_a$ and
$Q^x_a$. Our case is more special -- since there are no
hypermultiplets in 5 dimensions, one of the $SU(2)$ does not act on the bosons
and hence $Q^x=0$.

The variations of the gravitini and hyperini (in a vanishing
fermionic background) are then given by
\bea
\delta \psi_\mu^{\alpha\dot \alpha} &=& (D_\mu  \epsilon_{\dot
\alpha \dot \beta}+ Q_a^{\dot x} \sigma^{\dot x}_{\dot \alpha \dot
\beta}
 \pa_\mu \varphi^a )\eta_{\alpha\dot \beta } \\
\delta \chi^{A\alpha} &=&
V_a^{A\dot\alpha} \,\eta_{\alpha\dot\alpha}
\eea
where $\eta_{\alpha\dot\alpha}$ is the supersymmetry parameter, and
$V_a^{A\dot\alpha}$ is the quaternionic viel-bein, related
to the metric $G_{ab}$ and the quaternionic-K\"ahler
forms $\Omega^{\dot x}_{ab}$
by
\be
G_{ab} = V_a^{A\dot\alpha} \eps_{\dot\alpha\dot\beta} \Sigma_{AB}
V_b^{B\dot\beta}\ ,\quad
\Omega^{\dot x}_{ab} =  V_a^{A\dot\alpha}
\sigma^{\dot x}_{\dot\alpha\dot\beta} \Sigma_{AB} V_b^{B\dot\beta}
\ee
where $\Sigma_{AB}$ is the $Sp(2n+2)$ invariant antisymmetric tensor.
The quaternionic viel-bein $V^{A\dot\alpha}$ is a $2\times (2n+2)$ matrix,
which was computed for the c-map metric \eqref{cmap} in \cite{Ferrara:1989ik}:
\be
\label{quatviel}
V^{A\dot\alpha} = \begin{pmatrix}
\bar u &   v \\ -\bar  e^{\bar \I} &  E_{\I} \\ \bar E_{\bar \I}
& e^{\I} \\ - \bar v & u
\end{pmatrix}\ ,\quad
\ee
where
\bse
\bea
u &=& e^{-U + K/2} \,X^\Lambda \,
\left(d\tzeta_\Lambda  - {\cal N}_{\Lambda\Sigma} d\zeta^\Sigma \right) \\
v &=& dU + \frac{i}{2} \,e^{-2U} (d\sigma
+ \zeta^\Lambda d\tzeta_\Lambda - \tzeta_\Lambda d\zeta^\Lambda) \\
E_{\I} &=& i\, e^{-U} e_{\I}^i \, f_{i}^\Lambda 
 \left(d\tzeta_\Lambda  - \bar {\cal N}_{\Lambda\Sigma}
d\zeta^\Sigma \right)
\eea
\ese 
where $e^{\I}=e_{i}^\I dz^i$ is the holomorphic
viel-bein of the special K\"ahler manifold $\mathcal{M}_4$, 
such that $g_{i\bar j}=e_{i}^\I \bar e_{\bar j}^{\bar \J} \delta_{\I\bar \J}$,
and $e_\I^i$ is the inverse of $e_i^\I$.

\subsubsection*{The gauged case - some general formulas}

In the presence of gauging in 5 dimensions, the $\cN=4$ sigma model
in three dimensions has to be gauged. Our main goal in 
the rest of this subsection is to identify the appropriate gauging 
that corresponds
to the potential \eqref{V3}\footnote{We are grateful to H. Samtleben
for invaluable advice about gaugings of three-dimensional supergravities.}.

There are different equivalent ways of writing gauged sigma models
in three dimensions. While all the massless dynamical bosonic degrees of 
freedom have already been accounted for in the reduction, it
is possible to add auxiliary gauge fields which have no Maxwell-type
kinetic terms, but have Chern-Simons terms (in addition to their couplings 
to matter). In fact, in three dimensions a single gauge field
with a standard Maxwell kinetic term may be replaced by two gauge fields 
and a scalar~\cite{Nicolai:2003bp}. The two gauge fields are non-dynamical, 
and coupled by a Chern-Simons interaction. 
If there are no massless charged fields in the original Lagrangian, the two new
gauge fields can be integrated out, and the remaining 
scalar is just the dual of the original Maxwell field.
Otherwise, they need to be kept in the action since these two gauge fields still couple to 
matter fields via a $A_\mu J^\mu$ coupling which cannot be dualized. 
Note that in this formalism, vector fields always appear in pairs.

There are two sources of gauging and Chern-Simons vector fields in
3D. One is the gauging in 5D, which generates a potential in 5D
and hence a potential in 3D. The other source is the dependence
of the fermionic fields on the fibers in the reduction from 5D to 3D. 
In order to make the physical degrees of freedom apparent,
our aim is to dualize all vector fields into scalars.
This includes off-diagonal components of the 5D metric, which become
vector fields in 3D. The latter may be dualized into scalars 
provided there are no field charged under these gauge field. 
The bosonic fields are always invariant under translation in the fiber 
directions, hence are never charged. In contrast, the 3D fermions 
may or may not be constant in the directions of the fiber. If they are 
not constant then the fermions are charged under these vector fields and 
the sigma model will be inherently gauged. Note that by itself, 
this second source of gauging never generates a potential in 3D, 
since only the behavior of the fermion is affected. It is therefore
akin to the ``no scale supergravity models'' familiar in higher-dimensional 
supergravity \cite{Cremmer:1983bf,Gunaydin:1984ak}.

Both sources of the gauging, provided they preserve $\cN=4$ supersymmetry, 
may be described in the same formalism \cite{de Wit:2003ja,deWit:2004yr}. 
The global symmetries acting on the fields in the sigma model are
$G_3\times SU(2)_L\times SU(2)_R$, where $G_3$ is the isometry group
of the manifold $\cM_3$, and $SU(2)_L\times SU(2)_R$ is the R-symmetry
group. Gaugings are classified by a symmetric tensor
$\Theta_{\cM\cN}$, which encodes the embedding of the gauged symmetry
group into the global symmetry group. Here $\cM$ and $\cN$ live in the Lie
algebra of symmetries. The fact that $\Theta$ is symmetric is
related to the pairing of Chern-Simons vector fields discussed above.
We shall denote by $T^\cM=\{T^m, S^{\dot x}, S^x\}$, $m\
(n,p,..)=1,\dots \dim G_3, x\ (y,z)=1,2,3, {\dot x}\ ({\dot y},{\dot
z})=1,2,3 $ a basis of this Lie algebra. In the case (of interest in 
this paper) where only the
mixed components $\Theta_{m x}$ are non-zero, 
the condition that $\Theta_{\cM\cN}$ be invariant under the
gauge group (Eq. 3.13 in \cite{de Wit:2003ja}) reduces to
\be
\epsilon^{xyz}\,\Theta_{m x} \, \Theta_{n y} = 0\ ,\quad
f^{mn}_{\,p}\,\Theta_{m x} \,  \Theta_{n y} = 0\ .
\ee
These conditions are equivalent to the requirement that
$\Theta$ be decomposable,
\be
\Theta_{m x} = V_m\,n_x\ ,
\ee
where $n_x$ is a vector in $\IR^3$ and $V_m T^m$ an element in
$\mathfrak{g}_3$. Thus, this choice of projection tensor corresponds
to a rank 2 abelian gauge group $U(1)\times U(1)$, corresponding to
the generators $V_m T^m$ in $G_3$ and $n_x S^x$ in $SU(2)_R$,
respectively. Accordingly, one should introduce two Abelian gauge
fields $A_\mu$ and $B_\mu$ in three dimensions, with Chern-Simons
coupling $A dB$.

The other ingredient is the moment map
$\mathcal{V}^{\cM;[IJ]}=\cX^{\cM,a} Q_a^{[IJ]}+ S^{\cM,[IJ]}$.
Here, $\cX^{\cM,a} \pa_{\varphi^a}$ is the vector field on the
quaternionic-K\"ahler space $\cM_3$ corresponding to the symmetry
generator $T^\cM$, $Q_a^{[IJ]} d\varphi^a$ is the $SO(4)$ connection
on $\cM_3$, and $S^{\cM,[IJ]}$ is the compensating R-symmetry
induced by the action of  $T^\cM$. Rewriting the antisymmetric pair
of indices $[IJ]$ as either $x$ or $\dot x$, it is clear that
\be
\label{propv} \mathcal{V}^{x;y} = \delta^{xy}\ ,\quad
\mathcal{V}^{x;\dot x} = 0\ ,\quad \mathcal{V}^{m;x} = 0\ ,\quad
\ee
while $\mathcal{V}^{m;\dot x}$ is the usual moment map of
quaternionic isometries, defined by~\cite{Galicki:1986ja}
\be
d \mathcal{V}^{m;\dot x} + \epsilon^{\dot x \dot y \dot z} Q^{\dot
y} \wedge \mathcal{V}^{m;\dot z} = \cX^m \cdot \Omega^{\dot x}
\ee
where $\Omega^{\dot x}$ is the triplet of quaternionic two-forms
and $Q^{\dot x}$ is the $SU(2)_L$ connection.

With the embedding tensor $\Theta_{\cM\cN}$ and the moment map
$\mathcal{V}^{\cM;[IJ]}$ at hand, one can construct the $T$-tensors
\be
T^{IJ,KL}=\mathcal{V}^{\cM;[IJ]} \, \Theta_{\cM\cN}\,
\mathcal{V}^{\cN;[IJ]} \ ,\quad T^{IJ,a} = \mathcal{V}^{\cM;[IJ]} \,
\Theta_{\cM\cN}\, \mathcal{V}^{\cN;a}
\ee
where $\mathcal{V}^{\cM;a}\equiv \cX^{\cM,a}$, and the $A$-tensors
\be
A_1^{IJ}=\frac{1}{3} T^{MN,MN} \delta^{IJ} - 2 \, T^{IL,JL}\ ,\quad
A_{2,a}^{IJ}=\frac14 \left( D_a A_1^{IJ} + 2 T^{IJ}_{\,a} \right)
\ee
Using \eqref{propv}, we see that the only non-vanishing components of
the T-tensors are
\be
T^{x,\dot x} = \mathcal{V}^{\dot x}\, n^x \ ,\quad T^{x,a} = \cX^a\,
n^x\ ,\qquad \mathcal{V}^{\dot x} \equiv V_m\,\mathcal{V}^{m;\dot
x}\ ,\quad \cX^a \equiv V_m \cX^{m,a}
\ee
Therefore, the non-vanishing components of the $A$ tensors are
\be
A_1^{x\dot x}=\mathcal{V}^{\dot x}\, n^x\ ,\quad A_{2,a}^{x\dot
x}=\frac14 \cX^b \Omega_{ab}^{\dot x}\, n^x\ ,\quad A_{2,a}^{x} =
\cX_a\, n^x
\ee
Finally, the scalar potential is obtained from
\be
\label{V3gen} V_3 = -\frac{g^2}{4} \left( A_1^{IJ} A_1^{IJ} -2
g^{ab} A_{2,a}^{IJ} A_{2,b}^{IJ} \right) = -\frac{g^2}{4}
\left( \mathcal{V}^2 -\frac14 \cX^a g_{ab} \cX^b \right)
\ee

Using the formulas for $A_1$ and $A_2$ for this specific gauging,
and using Eq. 6.7 of \cite{deWit:2004yr}, we find that the
supersymmetry variation of the gravitini and the hyperini are:
\bea
\delta \psi_\mu^{\alpha\dot \alpha} &=& \left[ \mathcal{V}^{\dot x}
\sigma^{\dot x}_{\dot\alpha\dot\beta} \left( \epsilon_{\alpha\beta}
B_\mu +  \sigma^x_{\alpha\beta} u^x \gamma_\mu\right)+
\epsilon_{\dot\alpha\dot\beta} \, \sigma^x_{\alpha\beta} u^x
\, A_\mu\right]\,\eta_{\beta\dot\beta}\label{dpsil}\\
\delta \chi^{A\alpha} &=& V_a^{A\dot\alpha} \left[ \left( \pa_\mu
\phi^a + g \, \cX^a\, B_\mu \right) \gamma^\mu + g\, \cX^a
\right]\,\eta_{\alpha\dot\alpha} \label{dchil}
\eea
where $\eta_{\beta\dot\beta}$ is the supersymmetry parameter, and
$V_a^{A\dot\alpha} d\varphi^a$ is the quaternionic viel-bein
\eqref{quatviel}.

\subsubsection*{The gauging unmasked}

Our remaining task is to identify the Killing vector $\cX$ underlying the
scalar potential \eqref{V3}. The result depends on the source of gauging:

\begin{itemize}
\item[i)] {\it Gauging in 5 dimensions}: An important constraint is that
the Killing vector $\cX$ must commute with the action of 
the $Sl(2)$ symmetry exchanging
the $\pa_t$ and $\pa_\psi$ Killing vectors. Moreover, it should
commute with the Heisenberg generators. These constraints uniquely
determine
\be
\label{gauge} \cX = V_I \left( \pa_{\tzeta_I} - \zeta^I \pa_{\sigma}
\right),
\ee
where the $V_I$ are the coefficients that determine the F-I term and
the 5D potentianl in \eqref{V5}.

The $SU(2)$ connection on \eqref{cmap} being given by
\cite{Ferrara:1989ik}
\be
Q^{\dot 3} = e^{-2U} \left( d\sigma + \zeta^\Lambda d\tzeta_\Lambda
-\tzeta_\Lambda d\zeta^\Lambda \right)
+ \frac14 Q_K\ ,\quad Q^{\dot 1}= \Re(u)\ ,\quad Q^{\dot 2}=\Im(u)
\ee
where $Q_K=(\pa_{z^i} K \,dz^i - \pa_{{\bar z}^{\bar i}} K
d\bar z^{\bar i})/2i$ is the K\"ahler connection on $\mathcal{M}_4$,
we can compute the moment maps
\be
\mathcal{V}^{\dot 3} = e^{-2U}\, V_\Lambda\, \zeta^\Lambda\ ,\quad
\mathcal{V}^{\dot 1} = e^{-2U}\, V_\Lambda\, \Re(X^\Lambda)\ ,\quad
\mathcal{V}^{\dot 2} = e^{-2U}\, V_\Lambda\, \Im(X^\Lambda)\ .
\ee
Plugging into \eqref{V3gen}, and using   identities in \cite{Ceresole:2007rq},
we find that the scalar potential \eqref{V3gen} does indeed
reproduce \eqref{V3},
\be
V_3 = f^{-1} \,e^{-2U}\, V_5\ .
\ee

\item[ii)]{\it Gauging due to compactification}: 
The standard dimensional reduction from 5D to 3D assumes that the 
fermions are constant along the fibers. Some
of the solutions that we will discuss however, have a more
complicated variation along the fibers \cite{Sinha:2006sh}.
In such a case, the fermions are charged under the Maxwell 
gauge field associated to the off-diagonal metric 
in the fiber directions, which prevents its dualization 
into a pseudo-scalar. Thus, one obtains a pair of U(1) gauge fields
for every shift isometry under which the fermions  are charged. Since
the bosonic action is the same, irrespective of the charge of the
fermions, it had better be the case that this gauging does not change the
potential (and in particular, it does not introduce a potential if
there was none to start with).
Indeed, it is possible to check that the same potential is obtained for a 
one-parameter generalization of the Killing vector \eqref{gaugeX},
\be
\label{gaugeX}
\cX = V_I \left( \pa_{\tzeta_I} - \zeta^I \pa_{\sigma} \right)
+ \kappa \, \pa_\sigma
\ee
In order to convince oneself that $\pa_\sigma$ indeed the correct generator,
recall that the symmetries which shift the scalars
dual to the two Kaluza-Klein vector fields along $\pa_\psi$ and $\pa_t$ 
are $E_k$ and $E_{p_0}$, respectively. Thus the gauging has to be 
a linear combination of these two generators. The only one which
does not generate a potential is indeed $\cX=\kappa \pa_\sigma$.
\end{itemize}

\subsection{$\IR \times SU(2)\times U(1)$ symmetric solutions}

We now further specify to the case of stationary solutions with
an $SU(2)\times U(1)$ group of isometries. The metric on the three-dimensional
slices can be taken to be
\be
\label{ds3an}
ds_3^2 = N^2(\rho)\, d\rho^2 +
r^2(\rho)\, (\sigma_1^2 + \sigma_2^2)\ ,
\ee
where the lapse function $N$ maintains reparameterization invariance
along the radial direction $\rho$.
Here, $\sigma_i$ are the left-invariant $SU(2)$ forms\footnote{Note
the exchange of $\phi$ and $\psi$ with respect to \cite{Gutowski:2004yv}.},
\bea
\label{sutwoinv0}
\sigma_1&=& \sin\psi\, d\theta -\cos\psi\sin\theta\, d\phi\\
\sigma_2&=&\cos\psi\, d\theta+\sin\psi\sin\theta\, d\phi\\
\sigma_3&=& d\psi+\cos\theta\, d \phi
\eea
$\theta,\phi,\psi$ are the Euler angles of $S^3$ with ranges
$\theta\in[0,\pi)$, $\phi\in[0,2\pi)$, $\psi\in[0,4\pi)$, such that
\be
d\sigma^i= -\frac12 \epsilon_{ijk}\, \sigma^j \wedge \sigma^k
\ee
Moreover the coordinates $\varphi^a$ on $\cM_3$
are taken to be functions of $\rho$ only. It will be useful to relax
the condition that the $\psi$ circle has unit Chern class over the
$S^2$ parametrized by $(\theta,\phi)$, and define
\be
\sigma_{3,k}= d\psi+i k \cos\theta\, d \phi
\ee
For $k=-i$, the Hopf fiber $U(1)$ combines with $S^2$ to
produce the topology of $S^3$.

Upon reduction along the $\theta$ and $\phi$ direction, the
Lagrangian
\be
\label{L1}
\mathcal{L}_1 = N^{-1}
\left[ (r')^2 - r^2 \,g_{ab}\, \varphi^{'a} \varphi^{'b} \right] + N\, V_1
\ee
describes, in a reparametrization invariant way,
the motion of a fiducial particle on the cone $\IR \times \cM_3$
in the presence of a potential
\be
V_1 =  1 + r^2 e^{-2U} f^{-1} V_5\ .
\ee
In particular, the equation of motion of $N$ enforces the Hamiltonian
constraint
\be
\label{wdw}
H_{\rm WDW} \equiv
\frac{N}{16}
\left( p_r^2 - \frac{1}{r^2} g^{ab}\, p_{\varphi^{a}}\,p_{\varphi^{b}} - V_1
\right)= 0\ .
\ee
In the ungauged case, this mass of the particle is therefore fixed to unity,
and the motion decouples between the cone direction and $\cM_3$.
In the gauged case, the mass is effectively position dependent,
with a correction proportional to the radius $r e^{-U}/\sqrt{f}$
of the two-sphere measured in the five-dimensional metric. Moreover,
the cone direction and $\cM_3$ no longer decouple.
In either case, the phase space of $\IR\times SU(2)\times U(1)$
symmetric solutions of 5D supergravity is given by the symplectic quotient
\be
T^*( \IR^+ \times \mathcal{M}_3)\, // \, \{ H_{\rm WDW} = 0\} \ ,\quad
\ee
of dimension 16.

\subsubsection*{Conserved charges and integrability}

Due to the isometries \eqref{isoqk} of $\cM_3$, there are many conserved
quantities which can be used to integrate the motion. The
conserved charges $k, p^\Lambda, q_\Lambda, T_I$ are sufficient to eliminate
the (derivatives of) $\sigma, \zeta^\Lambda, \tzeta_\Lambda, \phi_I$.
Physically, the conserved quantity associated to $T_I$ and $q'_0$
correspond to the electric charge and angular momentum in 5 dimensions,
whereas $p^\Lambda, q_\Lambda$ are dipole-type charges. The charge $k$
is the Chern class of the circle bundle generated by $\pa_\psi$ over
the two-sphere parameterized by $(\theta,\phi)$, and should be
fixed to $k=-i$ (after analytic continuation) in order that the
total space have the topology of $S^3$. These identifications
are to be contrasted with the ones relevant for describing
four-dimensional black holes \cite{Gunaydin:2005mx,Gunaydin:2007bg},
and are consistent with the ``4D/5D lift'' \cite{Gaiotto:2005gf}
as shown in Section \ref{flipsec} below.

In the case where $\cM_3$ is a symmetric space $G_3/K_3$
and in the absence of gauging, all
solutions can in fact be obtained by exponentiating the action of the
isometry group\footnote{This fact was used in \cite{Gaiotto:2007ag}
to produce explicit non-supersymmetric extremal solutions in $D=4$, 
$\cN=2$ very special supergravity with one modulus.}. 
For this purpose, it is useful
to parametrize $G_3/K_3$ by an element $g$ in the Iwasawa gauge, i.e.
in the $A_3 N_3$ part of the Iwasawa decomposition of $G_3=K_3 A_3 N_3$
into the product of the maximal compact $K_3$, abelian torus $A_3$
and nilpotent subgroup $N_3$.
The right-invariant metric is obtained from the non-compact
part $p$ of the right-invariant one-form $\theta=dg\cdot g^{-1}$
valued in the Lie algebra $\mathfrak{g}_3$ of $G_3$,
\be
ds^2= {\rm Tr}(p^2)\ ,\quad \theta = h + p\ .
\ee
When $G_3$ is represented by real matrices, the Cartan decomposition
$\theta=h+p$ is simply the decomposition into antisymmetric matrices
$h$ and symmetric matrices $p$.
A geodesic passing through the point $g_0$ at $\tau=0$ with initial velocity
$p_0$ is then given by $g(\tau)=k(\tau)\cdot e^{ p_0 \tau}\cdot g_0$
where $p_0$ is a non-compact element in $\mathfrak{g}_3$, 
$k(\tau)$ is the unique element of $K_3$ which brings $g(t)$ back
to the Iwasawa gauge, and $\tau$ is the affine parameter. 
The rotation $k(\tau)$ drops from the
product $M(\tau)=g^t(\tau)\cdot g(\tau)$, from which 
the coordinates on $G_3/K_3$
can be read off. This produces a solution of the Lagrangian \eqref{L1}
in the gauge $N=r^2$ (Conversely,
given a solution of \eqref{L1}, the affine geodesic parameter $\tau$
may be obtained by integrating $N(\rho) d\rho / r^2(\rho) = - d\tau$).
The remaining motion of $r(\rho)$ may be obtained by integrating
the Hamiltonian
constraint \eqref{wdw}, and depends only on $p_0^2$; in particular,
if $(p^0)^2=0$, $r=1/(\tau+\gamma)$ where $\gamma$ is an integration
constant.  The $\mathfrak{g}_3$-valued conserved charges inherited 
from the right action of $G_3$ are then given by
\be
Q = - dM\,M^{-1} = - g_0^{t}\, p_0\, g_0^{-t}
\ee
Supersymmetric solutions correspond to special restrictions on the
momentum $p_0$. In many cases, but not all, $Q$ is nilpotent, i.e.
$Q^n=0, Q^{n-1}\neq 0$ for some integer $n$. Again, in the presence
of a potential $V_3$, $G_3$ is broken to a solvable subgroup and
integrability is lost in general.

\subsubsection*{Relation to Gauntlett {\it et al.} classification}
Supersymmetric solutions of $D=5, \cN=1$ supergravity were classified
in \cite{Gauntlett:2002nw} and \cite{Gauntlett:2003fk} for the ungauged
and gauged case, respectively. A necessary condition in the gauged
case is that the four-dimensional metric $ds_4^2$ in the
square bracket of \eqref{ds5gr} has to be K\"ahler. In terms of the
$SU(2)\times U(1)$ symmetric ansatz used in \cite{Gauntlett:2002nw,
Gauntlett:2003fk,Gutowski:2004yv}, slightly generalized to include
an arbitrary lapse function $\cN(\rho)$ and Chern class $k$,
\be
\label{ds5gr}
ds^2_5 = - f^2 (dt+\omega_4)^2 + f^{-1}
\left[ \cN^2 d\rho^2+ a^2(\sigma_1^2+\sigma_2^2)+ b^2 \sigma_{3,k}^2 \right]
\ee
\be
\omega_4 =\Psi \sigma_{3,k}\ ,\quad
A^I = f\,h^I\,(dt+ \Psi \sigma_{3,k}) + U^I(\rho)\,\sigma_{3,k}\ ,
\ee
a sufficient condition for K\"ahlerity is given by \cite{Gutowski:2004ez}
\footnote{This condition is in fact more restrictive than  K\"ahlerity.
For example, it is obeyed for flat $\IR^4$ or Eguchi-Hanson, but not for 
Taub-NUT.\label{foonut}}
\be
\label{ckahl1}
i k b \cN - 2 a a' = 0\ ,
\ee
corresponding to a K\"ahler form $\omega_K=-d( a^2\sigma_{3,k})$.
In terms of the variables in our Ans\"atze \eqref{ds4an},\eqref{ds3an},
related to the ones above by
\be
\label{abtoUr}
a = e^{-U} r\ ,\quad b= e^{U}\ ,\quad \cN = N e^{-U}
\ ,\quad \Psi= 2 i \zeta^0\ ,\quad U^I = \sqrt{\frac32}\,\zeta^I\ ,
\ee
the condition \eqref{ckahl1} becomes
\be
\label{ckahl2}
r' - r U' - i k \frac{N e^{2U}}{2 r} = 0\ .
\ee
It should be stressed that conditions \eqref{ckahl1} or \eqref{ckahl2}
are independent of the gauge coupling $1/\ell$. As we shall see below,
some supersymmetric solutions of ungauged supergravity 
do not satisfy this condition (e.g.
the Taub-NUT black hole). It therefore appears that 
more branches of solutions open up in the ungauged case $\ell=\infty$.
It would be interesting to see if remnants of these branches
exist at finite $\ell$.

\section{Supersymmetric solutions in $D=5, \cN=1$ minimal supergravity}
In this section, we specialize to the case of minimal $\cN=1$ supergravity
in 5 dimensions, possibly gauged, and restrict to stationary solutions
with $SU(2)\times U(1)$ group of isometries.

\subsection{Geometry of $G_{2(2)}/SO(4)$}
After the three-step reduction process explained in the previous section,
one obtains a one-dimensional Lagrangian
\be
\label{lagg2r}
{\cal L}=\frac{1}{N}\left[ (r')^2 - r^2 \left( u \,\bar u +
v \, \bar v +
e \, \bar e +
E \, \bar E \right) \right]
+ N \left( 1 + \frac{6r^2 e^{-2U}}{\tau_2 \ell^2} \right)
\ee
corresponding to the motion of a particle on the symmetric space
$\mathcal{M}_3=G_{2(2)}/Sl(2)\times Sl(2)$, with metric
\be
\label{dsm3}
ds_{\mathcal{M}_3}^2 =  J_{AB}\, \epsilon_{{\dot\alpha}{\dot\beta}}
\, V^{{\dot\alpha}A} \, V^{{\dot\beta}B} = u \,\bar u + v \, \bar v +
e \, \bar e + E \, \bar E \ ,
\ee
and, when $\ell$ is finite, a position-dependent mass. Here,
$J^{AB}$ and $\epsilon_{{\dot\alpha}{\dot\beta}}$ are
antisymmetric forms in 2 and 4 dimensions, with the conventions
$\epsilon_{12}=J_{14}=J_{32}=1$, and $V^{A\dot\alpha}$
is the quaternionic viel-bein with entries~\cite{Gunaydin:2007qq}
\bse
\label{quatvielg2}
\bea
u &=& \frac{e^{-U}}{ 2\sqrt{2}\, \tau_2^{3/2}}
\left( d\tzeta_0 + \tau\, d\tzeta_1
+ 3\, \tau^2 \,d\zeta_1 - \tau^3 \,d\zeta_0 \right)\\
v &=& dU + \frac{i}{2} e^{-2U} (
d\sigma - \zeta_0 d\tzeta_0 -\zeta_1 d\tzeta_1
+  \tzeta_0 d\zeta_0 +\tzeta_1 d\zeta_1 )\\
e &=& \frac{i \sqrt{3}}{2\tau_2} d\tau  \\
E &=&
-\frac{e^{-U}}{2\sqrt{6}\,\tau_2^{3/2} }
\left( 3 d\tzeta_0 + d\tzeta_1\, (\bar\tau + 2\tau)
+3 \tau\, (2\bar\tau+\tau) \,d\zeta_1 -3 \bar\tau\,\tau^2\,
d\zeta_0 \right)
\eea
\ese
The viel-bein $V^{A\dot\alpha}$ corresponds to the projection of
the right-invariant form $\theta=dg\cdot g^{-1}$ on the non-compact part $p$
of the Lie algebra of $G_{2(2)}$, parameterized in the Iwasawa gauge
as in \cite{Gunaydin:2007qq},
\be
\label{iwa1}
g=
\tau_2^{-Y_0} \cdot e^{\sqrt{2} \tau_1 Y_+}
\cdot e^{-U H} \cdot e^{-\zeta^0 {E_{q_0}} + \tzeta_0
{E_{p^0}}} \cdot e^{-\sqrt{3} \zeta^1 {E_{q_1}}
+ \frac{\sqrt{3}}{3} \tzeta_1 {E_{p^1}}} \cdot
e^{\sigma E_k}
\ee
where  $\tau\equiv \phi^1+i f \equiv \tau_1+i\tau_2$.
The entries of $V^{A\dot\alpha}$ in \eqref{quatvielg2}
have been normalized in such a way that
that $S_-$  acts by left multiplication by the standard spin 3/2
raising operator $\scriptsize\begin{pmatrix} 0 & \sqrt{3} & 0& 0 \\
0& 0& 2 & 0 \\ 0 & 0 & 0 & \sqrt3 \\ 0 & 0& 0& 0 \end{pmatrix}$
on the matrix \eqref{quatviel},
while $J_+$ acts by right multiplication
with $\scriptsize\begin{pmatrix} 0  & 1 \\ 0 & 0\end{pmatrix}$.

The compact part of the right-invariant form $\theta$ provides the
$SU(2)\times SU(2)$-valued spin connection
\be
\label{spincon}
\begin{pmatrix}
{\underline J}_+ \\ {\underline J}_3 \\{\underline J}_-
\end{pmatrix}= -{1\over 2}
\begin{pmatrix}
u \\ \frac{1}{4i}(v-{\bar v})+\frac{i\sqrt{3}}{4}(e-{\bar e})
\\\bar u
\end{pmatrix},
\begin{pmatrix}
{\underline S}_+ \\ {\underline S}_3 \\{\underline S}_-
\end{pmatrix}= \frac{\sqrt{3}}{2}
\begin{pmatrix}
\bar E \\ \frac{i\sqrt{3}}{4} (v-{\bar v}) + \frac{i}{4} (e-{\bar e}) \\  E
\end{pmatrix}.
\ee
As indicated below \eqref{cmap},
we are taking $\tau_1, \zeta^0,\tzeta_1,\sigma$
to be purely imaginary, so that we can use the same expressions
as in \cite{Gunaydin:2007qq} which was taylored for the Riemannian space
$G_{2(2)}/SO(4)$.
Our notations for the components of the right-invariant form $\theta$,
as well as for the Killing vectors of the right-action to be discussed
presently, are summarized in Figure \ref{g2rootdiag}.

\EPSFIGURE{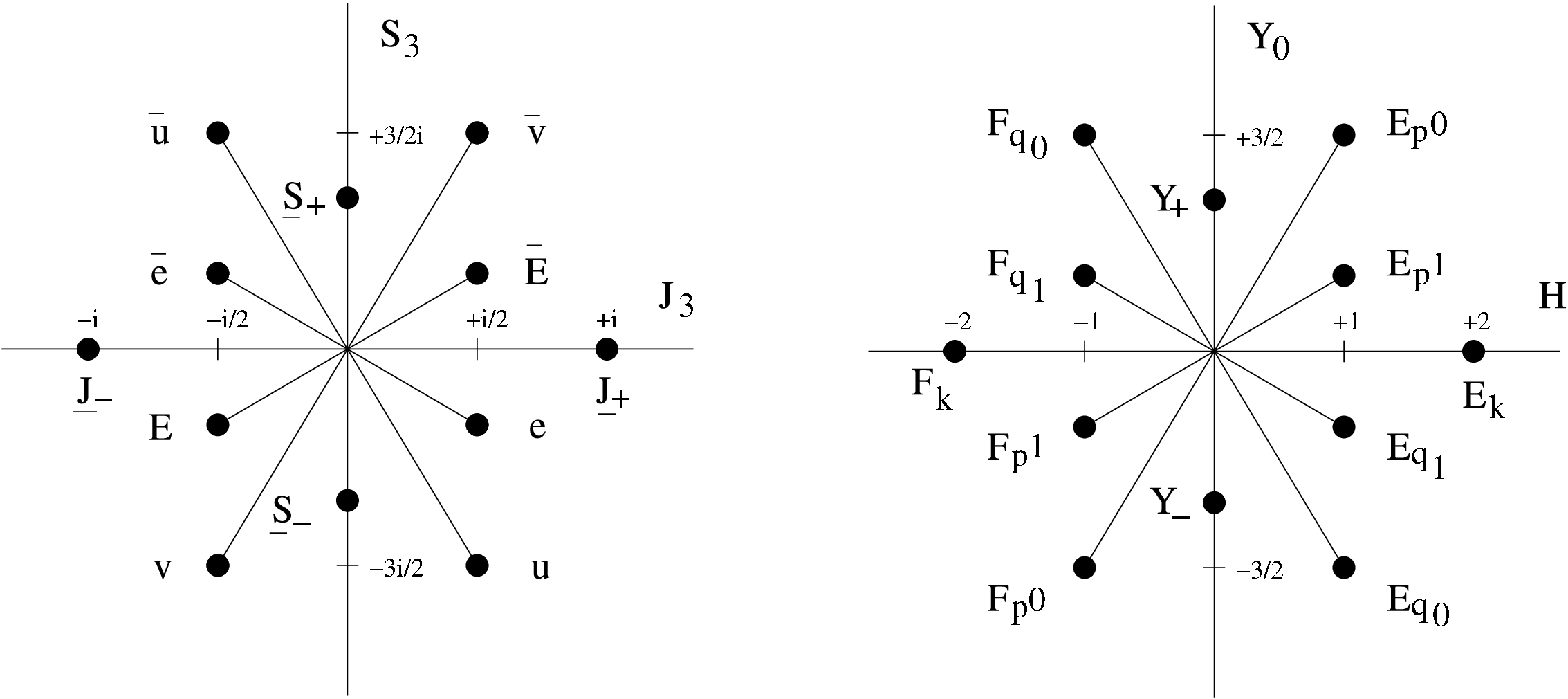, height=6.5cm}{
Left: components of the right-invariant
form $dg\cdot g^{-1}$, in the basis appropriate to the compact Cartan torus
$J_3,S_3$.
Right: Killing vectors for the right-action of
$G_{2(2)}$, in the basis appropriate
to the split Cartan torus $H,Y_0$.
\label{g2rootdiag}}

\subsubsection*{Conserved charges}

The metric \eqref{dsm3} on the symmetric space $G_{2(2)}/SO(4)$
is by construction invariant under the right-action of $G_{2(2)}$
on the coset representative $e$, compensated by a left-action of
its maximal compact subgroup such as to preserve the Iwasawa
gauge \eqref{iwa1}. The corresponding Killing vectors were
computed in \cite{Gunaydin:2007qq}. Replacing the vector field
$\pa_{\phi^a}$ by the momentum $p_{\phi^a}$ conjugate to $\phi^a$,
\be
p_U = 4 \frac{r^2}{N} U'\ ,\quad
p_{\tau_2} = 3  \frac{r^2}{N \tau_2^2} \tau_2'\ ,\quad
p_\sigma=\frac{r^2}{N} e^{-4U} (\sigma'+\tzeta_I  \zeta^{I'} - \zeta^I \tzeta_I')
\ ,\dots
\ee
we find that the conserved charges associated to the right-action
of  $G_{2(2)}$ are given by\footnote{For convenience, we stick
to the notations in \cite{Gunaydin:2007qq}. The generators 
$p^{\Lambda}, q^{\Lambda}, k, H, D, T_I, S^I, p_{'\Lambda}, q^{'\Lambda}, k'$
introduced in \eqref{isoqk} and subsequent equations in Section 2
are equal to $E_{p^\Lambda}, E_{q_\Lambda}, E_k, H$, $Y_0, Y_+, Y_-, 
F_{q^\Lambda}, F_{p_\Lambda}, F_k$, respectively. The lowest root $F_k$, too
bulky to be displayed here, can be obtained by Poisson commuting $F_{p^I}$
and $F_{q_I}$.}
\bse
\label{noetherc}
\bea
E_k&=&p_{\sigma}\\
E_{p^0}&=& p_{\tzeta_0} - \zeta^0 p_{\sigma} \quad\ ,\quad
E_{q_0}= -p_{\zeta^0} - \tzeta_0 p_{\sigma} \\
E_{p^1}&=& \sqrt{3}(p_{\tzeta_1} -  \zeta^1 p_{\sigma}) \quad\ ,\quad
E_{q_1}= \frac{1}{\sqrt{3}}(-p_{\zeta^1} -  \tzeta_1 p_{\sigma})\\
H &=& -p_U -2 \sigma p_{\sigma}
- \zeta^0 p_{\zeta^0} - \zeta^1 p_{\zeta^1}
-   \tzeta_0 p_{\tzeta_0}- \tzeta_1 p_{\tzeta_1}
\eea
\bea
Y_+&=& \frac{1}{\sqrt{2}}(p_{\tau_1} + \zeta^0 p_{\zeta^1}
- 6\zeta^1 p_{\tzeta_1} - \tzeta_1 p_{\tzeta_0})\\
Y_0&=&-\frac12(2\tau_1 p_{\tau_1} + 2\tau_2 p_{\tau_2}
- 3 \zeta^0 p_{\zeta^0}+ 3 \tzeta_0 p_{\tzeta_0}
- \zeta^1 p_{\zeta^1}+ \tzeta_1 p_{\tzeta_1})\\
Y_- &=& \frac{1}{3 \sqrt{2}} \left( 6 p_{\tau_2} {\tau_1} {\tau_2}+3 p_{\tau_1} \left(\tau_1^2-\tau_2^2\right)+9 p_{\tzeta_1} {\tzeta_0}-9
   p_{\zeta^0} {\zeta^1}+2 p_{\zeta^1} {\tzeta_1}\right)
\eea
\bea
F_{q_0} &=& -\frac1{\tau_2^3} \left(6 p_{\tzeta_1} (\zeta^1)^2-2 (p_{\tau_1}+p_{\zeta^1} {\zeta^0}) {\zeta^1}-p_{U} {\zeta^0}+2
   {\zeta^0} (p_{\tau_1} {\tau_1}+p_{\tau_2} {\tau_2}-p_{\zeta^0} {\zeta^0})
\right)\nn\\
&&\left. - p_{\tzeta_0} (\sigma+{\zeta^0}
   {\tzeta_0}+{\zeta^1} {\tzeta_1})+p_{\sigma} \left(2 (\zeta^1)^3+{\zeta^0} (-\sigma+{\zeta^0}
   {\tzeta_0}+{\zeta^1} {\tzeta_1})\right)\right) \\
&&+e^{2 U} \left(p_{\tzeta_0} \tau_1^3+p_{\sigma} {\zeta^0}
   \tau_1^3-3 p_{\tzeta_1} \tau_1^2-3 p_{\sigma} {\zeta^1} \tau_1^2+p_{\zeta^1} {\tau_1}-p_{\sigma} {\tzeta_1}
   {\tau_1}+p_{\zeta^0}-p_{\sigma} {\tzeta_0}\right) \nn
\eea
\bea
F_{q_1} &=& \frac{1}{3 \sqrt{3} \tau_2^3} \tau_2^3 \left(2 (p_{\tzeta_0}+p_{\sigma} {\zeta^0}) {\tzeta_1}^2+(-4 p_{\tau_1}-4 p_{\zeta^1} {\zeta^0}+3
   {\zeta^1} (5 p_{\tzeta_1}+p_{\sigma} {\zeta^1})) {\tzeta_1}\right.\nn\\
&&\left.-6 (2 p_{\tau_2} {\tau_1} {\tau_2}+p_{\tau_1}
   (\tau_1^2-\tau_2^2) ) {\zeta^0}-9 p_{\tzeta_1} (\sigma+{\zeta^0} {\tzeta_0})+9
   p_{U} {\zeta^1} \right.\nn\\
&&\left. +3 {\zeta^1} (-2 p_{\tau_1} {\tau_1}-2 p_{\tau_2} {\tau_2}+6 p_{\zeta^0} {\zeta^0}+3 p_{\sigma}
   (\sigma-{\zeta^0} {\tzeta_0})+2 p_{\zeta^1} {\zeta^1})\right)\nn\\
&&-3 e^{2 U} \left(3 p_{\zeta^0} {\tau_1}+p_{\zeta^1} \left(3
   \tau_1^2+\tau_2^2\right)+3
   \left(p_{\tzeta_0} {\tau_1} \left(\tau_1^2+\tau_2^2\right)-p_{\tzeta_1} \left(3 \tau_1^2+2
   \tau_2^2\right)\right.\right.\nn\\
&&\left.\left.+p_{\sigma} \left({\zeta^0} \tau_1^3-3 {\zeta^1} \tau_1^2+\tau_2^2 {\zeta^0}
   {\tau_1}-{\tzeta_0}-2 \tau_2^2 {\zeta^1}\right)\right) {\tau_1}-p_{\sigma} \left(3 \tau_1^2+\tau_2^2\right) {\tzeta_1}\right)
\eea
\bea
F_{p^0} &=& \frac{1}{27
   \tau_2^3}\left(2 p_{\sigma} {\tzeta_1}^3-6 p_{\zeta^1} {\tzeta_1}^2-9 (4 p_{\tau_2} {\tau_1} {\tau_2}+2 p_{\tau_1}
   (\tau_1^2-\tau_2^2)+6 p_{\tzeta_1} {\tzeta_0}-3 p_{\zeta^0} {\zeta^1}+3 p_{\sigma} {\tzeta_0}
   {\zeta^1}) {\tzeta_1}\right.\nn\\
&&\left.-27 (p_{\zeta^0} (\sigma-{\zeta^0} {\tzeta_0})+{\tzeta_0} (p_{U}+p_{\sigma} \sigma+2
   p_{\tau_1} {\tau_1}+2 p_{\tau_2} {\tau_2}+2 p_{\tzeta_0} {\tzeta_0}+p_{\sigma} {\zeta^0} {\tzeta_0}))\right)
   \tau_2^3\nn\\
&&+27 e^{2 U} \left(p_{\tzeta_0} \tau_1^6+p_{\sigma} {\zeta^0} \tau_1^6-3 p_{\tzeta_1} \tau_1^5-3 p_{\sigma}
   {\zeta^1} \tau_1^5+3 p_{\tzeta_0} \tau_2^2 \tau_1^4+3 p_{\sigma} \tau_2^2 {\zeta^0} \tau_1^4-6 p_{\tzeta_1}
   \tau_2^2 \tau_1^3\right.\nn\\
&&\left.+p_{\zeta^0} \tau_1^3-p_{\sigma} {\tzeta_0} \tau_1^3-6 p_{\sigma} \tau_2^2 {\zeta^1} \tau_1^3+3
   p_{\tzeta_0} \tau_2^4 \tau_1^2+p_{\zeta^1} \left(\tau_1^2+\tau_2^2\right) \tau_1^2+3 p_{\sigma} \tau_2^4
   {\zeta^0} \tau_1^2\right.\nn\\
&&\left.-p_{\sigma} \left(\tau_1^2+\tau_2^2\right) {\tzeta_1} \tau_1^2-3 p_{\tzeta_1} \tau_2^4
   {\tau_1}-3 p_{\sigma} \tau_2^4 {\zeta^1} {\tau_1}+p_{\tzeta_0} \tau_2^6+p_{\sigma} \tau_2^6 {\zeta^0}\right)
\eea
\bea
F_{p^1}&=&\frac{1}{3 \sqrt{3} \tau_2^3} \left(-\left((2 p_{\tzeta_1}-p_{\sigma} {\zeta^1}) {\tzeta_1}^2+2 p_{\tau_1} {\tau_1} {\tzeta_1}+(3 p_{U}+2 p_{\tau_2}
   {\tau_2}+6 p_{\tzeta_0} {\tzeta_0}+3 p_{\sigma} (\sigma+{\zeta^0} {\tzeta_0})) {\tzeta_1}\right.\right.\nn\\
&&\left.\left.-6 p_{\tau_1} \left(-2
   {\zeta^1} \tau_1^2+{\tzeta_0}+2 \tau_2^2 {\zeta^1}\right)+6 {\zeta^1} (4 p_{\tau_2} {\tau_1} {\tau_2}+6
   p_{\tzeta_1} {\tzeta_0}-3 p_{\zeta^0} {\zeta^1}+3 p_{\sigma} {\tzeta_0} {\zeta^1})
\right.\right.\nn\\
&&\left.\left.+p_{\zeta^1} (3 \sigma-3
   {\zeta^0} {\tzeta_0}+5 {\zeta^1} {\tzeta_1})\right) \tau_2^3-3 e^{2 U} \left(3 p_{\zeta^0} \tau_1^2-p_{\sigma}
   \left(3 \tau_1^2+2 \tau_2^2\right) {\tzeta_1} {\tau_1}
\right.\right.\nn\\
&&\left.\left.
+p_{\zeta^1} \left(3 \tau_1^3+2 \tau_2^2
   {\tau_1}\right)+3 \left(-p_{\tzeta_1} \left(\tau_1^2+\tau_2^2\right) \left(3
   \tau_1^2+\tau_2^2\right)-p_{\sigma} \left(\tau_1^2+\tau_2^2\right) {\zeta^1} \left(3
   \tau_1^2+\tau_2^2\right)
\right.\right.\right.\nn\\
&&\left.\left.\left.
+{\tau_1} \left(\left(\tau_1^2+\tau_2^2\right)^2 (p_{\tzeta_0}+p_{\sigma} {\zeta^0})-p_{\sigma}
   {\tau_1} {\tzeta_0}\right)\right)\right)\right)
\eea
\ese
As we shall see presently, the conserved charges $Y_+$ and $F_{q^0}$ correspond
to the electric charge and angular momentum, respectively.

For the most part, we will be interested in purely electric solutions,
which satisfy $E_{p^0}=E_{p^1}=0,E_k=k$ and $Y_+=q$. In this case, we may
solve for the time derivatives of the corresponding scalars as follows:
\bea
\label{taudot}
\tau_1' &=& -2  e^{-2 U} \tau_2^3  {\zeta^0} ({\zeta^1}'-{\zeta^0}'\,\tau_1)
+2 \frac{k N \tau_2^2}{r^2}  (\zeta^1-{\tau_1} {\zeta^0})^2
+\frac{\sqrt{2}}{3} \frac{q N \tau_2^2}{r^2} \ ,\\
\tzeta_0'&=& -2 \zeta^{0'}
   \tau_1^3+3 \zeta^{1'} \tau_1^2+e^{2 U} \frac{k N}{r^2}
   {\tau_2} \left( 3 \tau_1^2 {\zeta^0} -3 \tau_1 \zeta^1+\tau_2^2
   {\zeta^0}\right)\\
\tzeta_1'&=& 3 \tau_1^2 \zeta^{0'}
   -6 {\tau_1} \zeta^{1'} +3 e^{2 U} \frac{k N}{r^2}
   {\tau_2} (\zeta^1-{\tau_1}
   {\zeta^0})\\
\sigma'&=& e^{4 U}
   \frac{k N}{r^2} + \zeta^0 \tzeta_0' + \zeta^1 \tzeta_1' -
\tzeta_0 \zeta^{0'} - \tzeta_1 \zeta^{1'}
\eea
Moreover, we focus mostly on solutions with $k=-i$, such that
the angular directions $(\theta,\phi,\psi)$ parametrize a topological $S^3$.

\subsubsection*{4D-5D lift and $(t,\psi)$ flip\label{flipsec}}
It is also useful to introduce a different parametrization, adapted to the
$Sl(2,\IR)$ subgroup corresponding to the diffeomorphisms of the $(t,\psi)$
torus:
\be
g=V^{-\frac14(3 H+2 Y_0)}\,\cdot
\rho_2^{-\frac14(H-2 Y_0)}\,\cdot
e^{\frac{\sqrt2}{2} E_{q_0} \rho_1}\,\cdot
e^{\mu_1 E_{q_1}+ \frac{2}{\sqrt2} \mu_2 Y_+}\,\cdot
e^{\nu E_{p^1}}\,\cdot
e^{-\sqrt2 \tilde\mu_1 E_k + \tilde\mu_2 E_{p^0}}
\ee
The variables $(V,\rho_1,\rho_2,\mu_1,\mu_2,\nu, \tilde\mu_1,\tilde\mu_2)$
are related to the previous ones by
\be
V= e^U \sqrt{\tau_2},\quad
\rho_1=   -\sqrt{2} {\zeta^0}\ ,\quad
\rho_2= \frac{e^U}{\tau_2^{3/2}}\,\quad
\mu_1=\sqrt{3} ({\tau_1} {\zeta^0}-{\zeta^1})\ ,\quad
\mu_2=\sqrt{\frac{3}{2}} {\tau_1}\ ,\quad
\ee
\be
\nu = \frac{3 {\tau_1}   {\zeta^1}+{\tzeta_1}}{\sqrt{3}}\ ,\quad
\tilde \mu_1= -\frac{1}{\sqrt{2}} 
\left( \sigma+ \zeta^0 \tzeta_0 + \zeta^1\tzeta_1
-2 \tau_1^2 \zeta^0 \zeta^1 + 4 \tau_1 (\zeta^1)^2 \right)\,\quad
\tilde\mu_2= {\tzeta_0}-\tau_1^2 {\zeta^1},\quad
\ee
The parameter $\rho=\rho_1+i\rho_2$ transforms under this $Sl(2)$
by fractional linear transformations, while $(\mu_1,\mu_2)$ and
$(\tilde\mu_1,\tilde\mu_2)$ transform as doublets, and $\nu$ is inert.
In particular, under the Weyl reflection
\be
\rho\to -1/\rho\ ,\quad (\mu_1,\mu_2)\to  (\mu_2,-\mu_1)\ ,\quad
\nu\to\nu\ ,\quad (\tilde\mu_1,\tilde\mu_2)\to  (-\tilde\mu_2,\tilde\mu_1)
\ee
the coordinates $U,\tau,\zeta^I,\tzeta_I$ transform into
\be
\label{tpsiflip}
e^U \to \frac{e^U\,{\tau_2}^{3/4}}
{\left(2 (\zeta^0)^2 \tau_2^3+e^{2 U}\right)^{1/4}}\ ,\quad
\tau_1 \to \sqrt{2} ({\zeta^1}-{\tau_1} {\zeta^0})\ ,\quad
\tau_2 \to \sqrt{\frac{2 (\zeta^0)^2   \tau_2^3+e^{2 U}}{{\tau_2}}}\ ,\quad
\ee
\be
\zeta^0 \to -\frac{{\zeta^0} \tau_2^3}
{2 (\zeta^0)^2 \tau_2^3+e^{2 U}}\ ,\quad
\zeta^1\to -\frac{2 {\zeta^0} {\zeta^1} \tau_2^3+e^{2 U} {\tau_1}}
{\sqrt{2} \left(2 (\zeta^0)^2 \tau_2^3+e^{2 U}\right)}\ ,\quad
\ee
\be
\tzeta_0\to \frac{1}{\sqrt2} \left( -2
   (\zeta^0)^2 \tau_1^3+6 {\zeta^0} {\zeta^1} \tau_1^2-6 (\zeta^1)^2 {\tau_1}+\frac{4 \tau_2^3
   {\zeta^0} ({\tau_1} {\zeta^0}-{\zeta^1})^3}{2 (\zeta^0)^2 \tau_2^3+e^{2 U}}-\sigma-{\zeta^0}
   {\tzeta_0}-{\zeta^1} {\tzeta_1}\right)\ ,
\ee
\be
\tzeta_1 \to \frac{6 {\zeta^0}
   ({\zeta^1}-{\tau_1} {\zeta^0})^2 \tau_2^3}{2 (\zeta^0)^2 \tau_2^3+e^{2 U}}-3 \tau_1^2
   {\zeta^0}+6 {\tau_1} {\zeta^1}+{\tzeta_1}\ ,\quad
\ee
\be
\sigma \to \frac{{\zeta^0} (-\sigma+3 {\zeta^0} {\tzeta_0}+{\zeta^1} {\tzeta_1}) \tau_2^3+e^{2 U}
   \left({\zeta^0} \tau_1^3+{\tzeta_1} {\tau_1}+2 {\tzeta_0}\right)}{\sqrt{2} \left(2 (\zeta^0)^2
   \tau_2^3+e^{2 U}\right)}\ ,
\ee
Note that this transformation maps the reality conditions \eqref{rightcont}
appropriate to 5D black holes, to the conditions
to \eqref{othercont} appropriate to 4D black holes.
As we shall see on an example in the next section, it implements
the 4D/5D lift found in \cite{Gaiotto:2005gf}.

\subsubsection*{Poisson algebra of the viel-bein components}

In order to describe the constraints from unbroken supersymmetry, it will be
useful to compute the Poisson brackets of the entries in the quaternionic
viel-bein\footnote{Some of the results in this subsection were obtained in
collaboration with A. Waldron~\cite{piowal}.}.
By this we mean the following: the entries in
$V^{A\dot\alpha}$ are one-forms on $\cM_3$, which may be pulled back
to the world-line
into one-forms $v_a(\varphi) {\varphi^{a}}' d\tau$; expressing the velocities
${\varphi^{a}}'$ in terms of the momenta $p_{a}$, we obtain functions on the
phase space $T^*(\IR^+\times \cM_3)$, whose Poisson bracket can be computed
in the usual way. Equivalently, the one-form $v_a(\varphi)$ may be turned
into a vector field $\cX^a(\varphi)$ using the metric on $\IR^+\times \cM_3$,
and the Poisson bracket of $V$ is just the Lie bracket of $\cX$. The
result of this computation is that the non-vanishing Poisson brackets
among the entries of  $V^{A\dot\alpha}$, up to complex
conjugation, are given by
\bse
\label{poisbra}
\bea
\{v,\bar v\}=\{E,\bar E\}=-\{u,\bar u\}=\frac{N}{2 r^2}(\bar v-v)\ ,&\quad&
\{e,\bar e\}=\frac{N}{2\sqrt{3} r^2}(e-\bar e)\\
\{u,v\}=\{u,\bar v\}=-\frac{N}{4 r^2} u\ ,&\quad&
\{E,v\}=\{E,\bar v\}=-\frac{N}{4 r^2} E\\
\{u,e\}=-\frac{N\sqrt{3}}{4r^2}\,u\ ,&\quad&
\{u,\bar e\}=-\frac{N}{2r^2}\left(E-\frac{\sqrt{3}}{2}u\right)\\
\{E, e\}=-\frac{N}{2r^2}\left(u +\frac{\sqrt3}{6}\,E\right)\ ,&\quad&
\{\bar E,e\}=\frac{ N}{4\sqrt{3} r^2}(\bar E+4 E)
\eea
\ese
A useful observation is that, up to terms proportional to $V$ and $V'$,
the Poisson bracket $\{V,V'\}$ is given by a linear combination of the
spin connections $\utilde{J}_i$ and $\utilde{S}_i$ in \label{spincon},
with coefficients given by the Clebsh-Gordan coefficients for the
tensor product $(2,4)\wedge(2,4)=(1,3)+(3,1)+(1,7)+(3,5)$
in $SU(2)\times SU(2)$. In particular, the conservation of the
$J_3$ and $S_3$ charges may be checked easily from Figure
\eqref{g2rootdiag} (left).
The commutation relations \eqref{poisbra} can be summarized in the following
formula
\be
\label{VVJ}
\{V^{{\dot\alpha} A},V^{{\dot\beta} B}\} = {2i\over 3}
\epsilon^{{\dot\alpha}{\dot\beta}} {(\sigma_4^i)}^A_CJ^{CB}
{\underline S}_i + 2i
J^{AB}{(\sigma^i)}^{\dot\alpha}_{\dot\gamma}
\epsilon^{{\dot\gamma}{\dot\beta}}{\underline
J}_i + \dots
\ee
where $J^{AB}$ and $\epsilon_{{\dot\alpha}{\dot\beta}}$ are the
antisymmetric forms in 2 and 4 dimensions with the conventions
$\epsilon_{12}=J_{14}=J_{32}=1$. The dots in this expression denote
the $(1,7)$ and $(3,5)$ pieces in the tensor product, which vanish
when either $V^{{\dot\alpha} A}$ or $V^{{\dot\beta} B}$ vanish\footnote{A
more conceptual interpretation of the Poisson algebra \eqref{poisbra}
is that it is isomorphic to the Borel subalgebra of $G_{2(2)}$, after
an appropriate change of basis~\cite{piowal}.}.

Using \eqref{poisbra}, it is straightforward to compute the
commutation relations of the Hamiltonian
$H_0=u \,\bar u +v \, \bar v +e \, \bar e +E \, \bar E$
with the viel-bein:
\bse
\label{poisbraH}
\bea
\{H_0,u\} &=&\frac{N}{4 r^2} \left(u
\left(- \sqrt{3} (e- {\bar e}) +3 \bar v- v\right)+2  e E\right)\\
\{H_0,v\} &=& -\frac{N}{2
   r^2} \left(E {\bar E}+u {\bar u}+v ({\bar v}-v)\right)\\
\{H_0,e\} &=& \frac{N}{2 \sqrt{3} r^2}
\left(-e (e-{\bar e}) +2 E^2-2 \sqrt{3} \bar E   u\right)\\
\{H_0,E\} &=& \frac{N}{4 \sqrt{3} r^2}
\left(- E e-4  e \bar E+ \bar e \left(E+2
   \sqrt{3} u\right)+\sqrt{3} E (3 v-\bar v)\right)
\eea
\ese
These relations will become useful when analyzing the algebra
of constraints imposed by supersymmetry in the next Section.

\subsubsection*{Shifted quaternionic vielbein}
In view of the supersymmetry variation \eqref{dchil} in
the gauged case, we define the ``shifted quaternionic vielbein''
\be 
\utilde{V}^{A \dot \alpha} = V^{A \dot \alpha}_a\, \left( d\varphi^a -
\frac{3 i \sqrt2}{\ell} N\, \cX^a \right) \ ,
\ee 
where
$\cX=\pa_{\tzeta_1}-\zeta^1 \pa_\sigma$ is the Killing vector
controlling the gauging, and the normalization has been chosen to
agree with the analysis in \cite{Gutowski:2004ez}. In terms of the
entries of the ``shifted vielbein'',
\be \label{vbeint}
\utilde{u}=u+ \frac{3 e^{-U} N}{2\ell \tau_2^{3/2}} (\tau_2-i
\tau_1)\ ,\quad \utilde{v}=v+\frac{3\sqrt2}{\ell} N e^{-2U} \zeta^1\
,\quad \utilde{E}=E+\frac{\sqrt3 e^{-U} N}{2\ell \tau_2^{3/2}}
(\tau_2-3 i \tau_1)\ ,\quad \utilde{e}=e
\ee
Note that the deformation does not commute with complex conjugation:
\be
\label{vbeintb}
\utilde{\bar u}=\bar u- \frac{3 e^{-U} N}{2\ell \tau_2^{3/2}}
(\tau_2+i \tau_1)\ ,\quad \utilde{\bar v}=\bar
v-\frac{3\sqrt2}{\ell} N e^{-2U} \zeta^1\ ,\quad \utilde{\bar
E}=\bar E-\frac{\sqrt3 e^{-U} N}{2\ell \tau_2^{3/2}} (\tau_2+3 i
\tau_1)\ ,\quad \utilde{\bar e}=\bar e \nn
\ee

\subsection{The BMPV and Taub-NUT black holes}
In the next three sections, we various reductions of the
one-dimensional dynamical system relevant
for different kind of black holes.

\subsubsection*{Constraint analysis}
In the absence of gauging, the one-dimensional system is the same as
the one which arises in the study of four-dimensional
black holes~\cite{Gunaydin:2005mx,Gunaydin:2007bg,Neitzke:2007ke}.
In this case, the supersymmetry conditions are given by
\be
\label{veps}
r'=N\ \qquad \mbox{and} \qquad \exists\, \epsilon_{\dot\alpha} \quad \mbox{\rm such that} \quad
V^{A \dot\alpha} \epsilon_{\dot\alpha} = 0\ .
\ee
The second condition is equivalent to the vanishing of the $2\times 2$
minors of the quaternionic vielbein,
\be
H_{AB} \equiv V^{A \dot\alpha} V^{B \dot\beta} \epsilon_{\dot\alpha
\dot\beta} = 0\ .
\ee
In particular, it implies the vanishing of the total momentum
$P^2=H_{AB} \Sigma^{AB}$ on $\mathcal{M}_3$, and together with
$r'=N$ the vanishing of the total Hamiltonian \eqref{wdw}.

Using \eqref{poisbra}, one may check that the Poisson bracket of $H_{\rm WDW}$
with the constraints $H_{AB}$ vanishes on the locus $H_{AB}$, for example
\be
\{H_{12},H_{13}\} =
-\frac{\sqrt{3}}{3}  u\, H_{12}+\frac{\sqrt{3}}{6} \bar v\, H_{13}
- \frac{1}{2} e \, H_{13} = 0\ .
\ee
This implies that the projectivized
Killing spinor $z=\epsilon_{\dot 2}/\epsilon_{\dot 1}$ can be computed
consistently from any of the equations $z=-V^{A\dot 1}/V^{A\dot 2}$;
using again \eqref{poisbra}, one may check that
\be
\label{dzP}
z' - \frac12 \left[ \bar u + \frac12
[(v-\bar v) - \sqrt3 (e-\bar e)] z +  u\, z^2\right] = 0Ê\ .
\ee
Using the results in \cite{Gunaydin:2007qq}, one may check that this
is precisely the condition $(dz+P)/d\rho=0$, where $P$ is the projectivized
$SU(2)$ connection on the quaternionic-K\"ahler manifold $\mathcal{M}_3$.
Therefore, the motion can be lifted to a holomorphic geodesic on the twistor
space $Z=G_{2(2)}/SU(2)\times U(1)$ \cite{Gunaydin:2007bg,Neitzke:2007ke}.

\subsubsection*{The generalized BMPV black hole}
It is straightforward to check that the constraint \eqref{veps}
indeed describes the supersymmetric spinning
black hole constructed in \cite{Breckenridge:1996is}. In fact, the
BMPV black hole is part of a family of solutions given
by \cite{Gauntlett:2002nw}
\be
\label{bmpv}
\cN=1\ ,\quad a=b=\frac{\rho}{2}\ ,\quad
\tau_2 = -i \tau_1 =
f = \left(\lambda + \frac{\mu}{\rho^2} + \frac{\chi^2}{9} \rho^2\right)^{-1}
\ ,\quad
\ee
\be
\Psi = \frac{j}{2 \rho^2} + \frac{\chi\mu}{2} + \frac{\chi\lambda}{4}
\rho^2 + \frac{\chi^3}{54} \rho^4\ ,\quad
U^1=\frac1{4\sqrt3} \chi \rho^2\ ,\quad
\ee
where $\mu$ is the electric charge, equal to the ADM mass by the BPS
condition, $j$ is the angular momentum, and $\chi$ is a deformation parameter
which does not preserve the asymptotic flatness.
Using \eqref{taudot}, we easily obtain the remaining coordinates of the
non-linear sigma model,
\be
\sigma = -i \frac{\rho^2}{4} + i \frac{j \tzeta_0}{2\sqrt{2} \rho^2}
+ i\frac{\lambda \chi \tzeta_0 \rho^2}{4\sqrt2}
-\frac{\chi \tzeta_1 \rho^2}{6\sqrt2}
+i \frac{\chi^3 \rho^4 \tzeta_0}{54\sqrt2}
\ee
while $\tzeta_0$ and $\tzeta_1$ take constant values. It is easy
to check that the first column\footnote{This could be traded
for the second column upon flipping the sign
of $\tau_1,\Psi,\tzeta_1,\sigma$, corresponding to a flip
of $i$ in \eqref{rightcont}.} of the viel-bein $V$ vanishes\footnote{Moreover,
we observe that  $E=0$ when $\chi=0$.}:
\be
\label{ueEv}
\bar u = \bar e = \bar E = \bar v = 0
\ee
so that the supersymmetry constraints \eqref{veps} are obeyed at $z=0$,
which is indeed a fixed point of \eqref{dzP} when $\bar u=0$. It would
be interesting to find a more general class of solutions where an
arbitrary linear combination of the two columns of the quaternionic
viel-bein vanishes. At any rate, the fact that the solution preserves
the same supersymmetry condition as the one appropriate for 4D black
holes is consistent with the fact that the Killing spinor
preserved by the 5D solution is independent of the $\psi$
direction~\cite{Gauntlett:1998fz}.

It is also instructive to compute the conserved charges \eqref{noetherc}
for the generalized BMPV solution:
\be
E_{p^0}=E_{p^1}=0\ ,\quad
E_{q_0}=2i\tzeta_0, \quad E_{q_1}= \frac{2i}{\sqrt3} \tzeta_1\ ,
\quad E_k=-i\ ,\quad
\ee
\be
Y_0=0\ ,\quad
Y_+ = \frac{3i\mu}{4\sqrt2}\ ,\quad
Y_-=-\frac{i\sqrt 2 \tzeta_1^2}{3}\ ,\quad
H=\frac{\mu \chi \tzeta_0}{\sqrt2}\ ,\quad
\ee
\be
F_{p^0} = \frac{\chi \mu \tzeta_0^2}{\sqrt2} + \frac{4 i \tzeta_1^3}{27}\ ,
F_{p^1}=\frac{\mu \tzeta_0 ( 3 i + \sqrt2 \chi \tzeta_1)}{2\sqrt3}\ ,\quad
F_{q_1}=-\frac{i \mu \tzeta_1}{\sqrt3}\ ,\quad
F_{q_0}=-\frac{i j}{2\sqrt2}\ ,\quad
\ee
\be
F_k = -\frac{i}{24}\left[ 6 \sqrt2 j \tzeta_0 + \mu
\left( 3 \chi^2 \mu \tzeta_0^2 + 4 \tzeta_1^2 \right) \right]\ ,\quad
\ee
This confirms the identification of $Y_+$ and $F_{q_0}$ as the
electric charge and angular momentum, respectively.
Moreover, one may check that the full
Noether charge (viewed as a $7\times 7$ matrix via $G_{2(2)}\subset
SO(3,4)$, see  \cite{Gunaydin:2007qq}) is nilpotent of degree 3, $Q^3=0$.
This is a general consequence of the supersymmetry condition \eqref{veps}
in the symmetric space case~\cite{Gunaydin:2005mx}. However, it follows
more generally from requiring extremality: indeed, a smooth near-horizon
geometry is obtained provided $a^2/f$ and $b^2/f$ take a finite value
as $\rho\to 0$, and $f\sim\rho^2$ in the gauge $\cN=1$. In terms of
the affine geodesic parameter $\tau=4/\rho^2$, we see that the
entries in the matrix $M=e(\tau) e^t(\tau)$ (particularly
the entry $M_{44}=e^{-2U}/\tau_2=1/(b^2 f)$, in the basis
used in \cite{Gunaydin:2007qq}) grow at most like $\tau^2$,
consistently with $Q^3=0$.

\subsubsection*{The Taub-NUT black hole}
Another solution in the same category is the rotating BPS black hole at the tip
of a Taub-NUT space \cite{Gaiotto:2005gf}
\footnote{This may be obtained as a special case
$\lambda=1,\chi=0,\delta=\mu/\sqrt{L},
a=\sqrt{L/4}$ of the family of rotating Taub-NUT solutions
in \cite{Gauntlett:2002nw} (Eq. 3.57-59), upon changing coordinates
from $\rho$ to $r$ such that  $\rho=2 a ( r- a)$.}
with NUT charge $L$, electric charge $\mu$ and angular momentum $j$
\be
\label{tnsol}
\cN=\sqrt{1+ \frac{L}{\rho}}\ ,\quad
a = \rho \sqrt{1+ \frac{L}{\rho}}\ ,\quad
b = 1/\sqrt{1+ \frac{L}{\rho}}\ ,\quad
\tau_2=-i \tau_1 = f = \left( 1 + \frac{\mu}{\rho} \right)^{-1}\ ,\quad
\ee
\be
\Psi = \frac{j}{L}\left(1 + \frac{L}{\rho}\right)\ ,\quad
\tzeta_0=\tzeta_1=0\ ,\quad \sigma = \frac{i L}{\rho+L}\ .
\ee
The only non-vanishing conserved charges are given by
\be
E_k=i L\ ,\quad
H=-2L\ ,\quad
Y_+ = \frac{3i\mu}{\sqrt2}\ ,\quad
F_{q_0}=-2i\sqrt2 j\ ,\quad
F_k = i L
\ee
while the affine geodesic parameter is $\tau=1/\rho$.
Just like the BMPV black hole \eqref{bmpv},
the conditions \eqref{ueEv} are obeyed and  $Q^3=0$ (as the two share
the same near horizon geometry).
In fact the two families of solutions are related by a
``D-transformation''
in the language of \cite{Giusto:2007fx,Ford:2007th}, i.e. a Weyl reflection
$D=\exp\left[\frac{i \pi}{4}(E_k + F_k) \right]$. Moreover, the solution
\eqref{tnsol} is related by a $(t,\psi)$ flip \eqref{tpsiflip}
to a 4D BPS black hole with Noether charges
\be
E_{p^0} = - i L \sqrt2\ ,\quad
E_{p^1} =0\ ,\quad
E_{q_1}=-i \mu \sqrt6\ ,\quad
E_{q_0}= 2i j \sqrt2\ ,\quad
E_k = 0\ .
\ee
This is consistent with the 4D/5D lift \cite{Gaiotto:2005gf}.
For the pure Taub-NUT vacuum, i.e. when $j=\mu=0$, we note that the matrix of
Noether charges becomes nilpotent of degree 2. In appendix \ref{nil2app},
we provide more general solutions of this type.

\subsubsection*{The semi-classical radial wave function of
BMPV-type black holes}
The five constraints
\be
\bar u = \bar e = \bar E = \bar v = p_r+4 = 0
\ee
are solved in the sector
\be
Y_+ = q\ ,\quad E_k = k
\ee
by setting $p_a=\pa_{\phi^a}S_{q,k,\mathcal{S}}$, where
\be
\begin{split}
S_{q,k, \mathcal{S}} =& -4 r + i k e^{2U} + \sqrt2 q \bar\tau
+ k \left[ \sigma + \zeta^0 \tzeta_0 + \zeta^1 \tzeta_1 +
6 \bar\tau \zeta_1^2 - 6 \bar\tau^2 \zeta^0\zeta^1
+2 \bar\tau^3 (\zeta^0)^2 \right]\\
&+ \mathcal{S} \left( \tzeta_0 + \bar\tau \tzeta_1 +3 \bar\tau^2 \zeta^1
- \bar\tau^3\zeta^0, \tzeta_1+ 6 \bar\tau \zeta^1
- 3 \bar\tau^2 \zeta^0 \right)
\end{split}
\ee
and $\mathcal{S}$ is an arbitrary function of two variables.
Thus, a basis of solutions of the quantum constraints
in the semi-classical approximation is given by
\be
\Psi_{q,k,p^0,p^1} \sim
\exp\left[ i S_{q,k,0} + i p^0
\left( \tzeta_0 + \bar\tau \tzeta_1 +3 \bar\tau^2 \zeta^1
- \bar\tau^3\zeta^0\right)
+ i p^1 \left( \tzeta_1+ 6 \bar\tau \zeta^1
- 3 \bar\tau^2 \zeta^0\right)
\right]
\ee
While $p^0$ is indeed the eigenvalue of $E_{p^0}$, $E_{p^1}$ and $F_{q_0}$
do not commute with $Y_+$, and so cannot be diagonalized simultaneously.
Note that the wave function is a anti-holomorphic function of $\tau$, as
required by the condition $\bar e=0$, and that it flattens out in
the near horizon region where $a,b\to 0$, in the case of 4D black
holes \cite{Neitzke:2007ke,Gunaydin:2007bg}.

\subsection{The G\"odel and Eguchi-Hanson black holes}
We now turn to a second class of supersymmetric solutions, which turn out
to satisfy a different set of constraints.

\subsubsection*{The G\"odel black hole}
In addition to the above families of solutions in $D=5$ minimal (ungauged)
supergravity, a second class of supersymmetric solutions with isometries
$SU(2)\times U(1)$ was constructed \cite{Gauntlett:2002nw}:
\bse
\label{godel}
\be
\cN=1\ ,\quad a=b=\frac{\rho}{2}\ ,\quad
\tau_2 = i \tau_1 = f = \left(\lambda+\frac{\mu}{\rho^2}
+\frac{\chi^2}{27\rho^6}\right)^{-1}\ ,\quad
\ee
\be
U^1 = \frac{\chi}{4\sqrt3 \rho^2}\ ,\quad
\Psi=\gamma\,\rho^2
-\chi \left( \frac{\lambda}{4\rho^2}+\frac{\mu}{6\rho^4}+
\frac{\chi^2}{270\,\rho^8} \right)
\ee
Note that the relative sign between $\tau_1$ and $i \tau_2$
is flipped with respect to the BMPV solution \eqref{bmpv}.
As above, $\tzeta_0$ and $\tzeta_1$ take constant values, while
\be
\sigma = -\frac{i}{4} \left( 1-2\sqrt2 \gamma \tzeta_0 \right)\rho^2
-\frac{\chi(2 \tzeta_1 + 3 i \lambda \tzeta_0)}{12\sqrt2 \rho^2}
-\frac{i \chi \mu \tzeta_0}{6\sqrt2 \rho^4}
 -\frac{i \chi^3 \tzeta_0}{270\sqrt2 \rho^8}
\ee
\ese
These solutions are deformations of the generalized G\"odel solution
obtained by setting $\mu=\chi=0$.
For these G\"odel-type solutions, the conserved charges are given by
\bse
\be
E_{p^1}=E_{p^0}=0\ ,\quad
E_{q_0}=2i\tzeta_0, \quad E_{q_1}= \frac{2i}{\sqrt3} \tzeta_1\ ,
\quad E_k=-i\ ,\\
\ee
\be
H=Y_0=0\ ,\quad
Y_+=-\frac{3 i \mu}{4\sqrt2}\ ,\quad
Y_-=-\frac{i\sqrt 2 \tzeta_1^2}{3}\ ,\quad
\ee
\be
F_{q_0}=\frac{i \chi \lambda}{4\sqrt2}\ ,\quad
F_{q_1}=\frac{1}{4\sqrt3}( \sqrt2 \chi + 4 i \mu \tzeta_1)\ ,\quad
F_{p^1}=-\frac{i}{2} \sqrt3 \mu \tzeta_0\ ,\quad
F_{p^0}=\frac{4 i \tzeta_1^3}{27}\ ,\quad
\ee
\be
F_k=\frac{i}{6} \mu \tzeta_1^2 + \frac{\sqrt2}{24} \chi
( 2 \tzeta_1+ 3 i \lambda \tzeta_0)
\ee
\ese
and the geodesic affine parameter is $\tau=4/\rho^2$.
Unlike the BMPV and Taub-NUT black holes, which had $Q^3=0$,
one may check that the Noether charge matrix is
now nilpotent of degree 7, $Q^7=0$,
This shows that these solutions do
not arise from the same supersymmetry constraint \eqref{veps}.
Indeed, one may check that three out of the four entries of the
{\it second column} of the quaternionic viel-bein vanishes,
\be
\label{eEu}
e = E =  u = 0 \ ,\quad
\ee
moreover
\be
\label{vnu}
C_1 \equiv i k v-N\,e^{-2U}=0\ ,\qquad
C_2 \equiv r' - r U' - i k \frac{N e^{2U}}{2 r}=0
\ee
and
\be
\label{c3c4}
C_3 \equiv \bar v = 0 \ ,\quad C_4 \equiv r^2 + k^2 e^{4U} = 0\ .
\ee
The condition $C_2=0$ is recognized as the K\"ahlerity
condition \eqref{ckahl2}.

\subsubsection*{The  Eguchi-Hanson black hole}
The conditions $C_3=C_4=0$ turn out to be relaxed in the case
of the more general Eguchi-Hanson black holes found
in \cite{Gauntlett:2002nw}:
\bse
\label{eguchi}
\be
\cN=\frac{1}{\sqrt{1-m^4/\rho^4}}\ ,
\quad a=\frac{\rho}{2}\ ,\quad
b=\frac{\rho}{2}\sqrt{1-m^4/\rho^4}\ ,\quad
\ee
\be
\tau_2 = i \tau_1 = f = \left[\lambda
-\frac{\chi^2}{9 m^4 \rho^2}
+\delta \log\left(\frac{\rho^2-m^2}{\rho^2+m^2}\right)
\right]^{-1}\ ,\quad
U^1 = \frac{\chi}{4\sqrt3 \rho^2}\ ,\quad
\ee
\be
\Psi=\gamma\,\rho^2
-\frac{\chi\lambda}{4\rho^2}+\frac{\chi^3}{54 m^4\rho^4}+
\frac{\delta\chi}{4\rho^2 m^4}
\left[ (\rho^4-m^4) \log\left(\frac{\rho^2-m^2}{\rho^2+m^2}\right)
+2 m^2 \rho^2\right]
\ee
\be
\tzeta_0 = {\rm cte}\ ,\quad \tzeta_1={\rm cte}\ ,\quad
\sigma = -\frac{i}{4} \rho^2 - \frac{i m^4}{4 \rho^2}
-\frac{\chi}{6\sqrt2 \rho^2} \tzeta_1
-\frac{i \chi\delta}{2\sqrt2 m^2} \tzeta_0 + \frac{i}{\sqrt{2}} \Psi \tzeta_0
\ee
\ese
where $\tzeta_0$ and $\tzeta_1$ are constants. The conserved
charges are given by
\bse
\be
E_{p^0}=E_{p^1}=0, \quad
E_{q_0}=2 i \tzeta_0,\quad
E_{q_1}=\frac{2 i \tzeta_1}{\sqrt3}\ ,\quad E_k=-i\ ,\quad Y_0=0\ ,
\ee
\be
H=\frac{\chi \delta \tzeta_0}{m^2 \sqrt 2}\ ,\quad
Y_+=\frac{1}{12\sqrt 2 m^4}(\chi^2+18 \delta m^6)\ ,\quad
F_{q_0}=\frac{i}{4\sqrt 2}(\chi \lambda-4 \gamma m^4)\ ,\quad
F_{p^0}=-\frac{i\sqrt2}{3}\tzeta_1^2\ ,\quad
\ee
\ese
while $F_{q_1},F_{p^1},F$ are too bulky to be displayed.
The affine geodesic parameter is related to $\rho$
by $\tau=4 {\rm arctanh}(\rho^2/m^2) / m^2$. For regularity
in the range $\rho\geq m$
one must impose $\delta=0$ and $\chi^2\leq 9\lambda m^6$.
Moreover closed time-like curves at $r\to\infty$ can be
avoided by taking $\gamma=0$. Note that this is also possible
to analytically continue $m\to e^{i\pi/4} m$, in such a way
that the solution is regular on the $\rho>0$ axis.

For these Eguchi-Hanson black holes, the conditions \eqref{eEu} and
\eqref{vnu} are satisfied, but $\bar v\neq 0$ and the
Noether charge matrix is no longer nilpotent. Instead, its
Jordan form in the $7\times 7$ matrix representation
has one $3\times 3$ nilpotent block
and two $2\times 2$ blocks of the form $\scriptsize
\begin{pmatrix} \pm m^2 & 1 \\
0 & \pm m^2 \end{pmatrix}$.

\subsubsection*{Constraint analysis}

This motivates us to study the reduction of the one-dimensional
dynamical system under the three constraints \eqref{eEu} (still
restricting to the ungauged case in this section).
Using \eqref{poisbra},\eqref{VVJ} and \eqref{poisbraH},
it is straightforward to check that
the three constraints \eqref{eEu} are first class, {\it i.e.}
that they Poisson-commute among each other and
with the Hamiltonian constraint \eqref{wdw}
on the constraint locus. Thus the Hamiltonian system
\eqref{wdw} admits a consistent reduction to
the 12-dimensional symplectic quotient
\be
\label{teEu}
T^*( \IR^+ \times\mathcal{M}_3) \,//\,
\{ e = E = u = 0\}\ .
\ee
On this locus, the Hamiltonian \eqref{wdw} reduces to
\be
H_{\rm WDW} = N \left( \frac{1}{16} p_r^2 -1 \right) -\frac{r^2}{N}v\bar v
=N \left[ \frac{1}{16} \left( p_r^2 - \frac{p_U^2}{r^2}\right) -
\frac{e^{4U} k^2}{4 r^2} -1 \right]\ .
\ee
Moreover, the generators $E_{p^I},E_{q_I},E_k,H,Y_0,Y_\pm$
also commute with these constraints, and therefore lead to an action of
$\IR\times G_4 \times H_5$ on the phase space \eqref{teEu}.

Having imposed the constraints \eqref{eEu}, it may
be checked easily that the additional constraint $v=0$,
 commutes with
$e=E=u=0$ as well as with $H_{WDW}$, thus proving the
consistency of the constraints \eqref{ueEv} (or their complex
conjugates) relevant for the BMPV and Taub-NUT black holes.

Instead, we want to enforce the constraints $C_1=C_2=0$ in \eqref{vnu},
which were found to govern the G\"odel and Eguchi-Hanson black holes.
Rewriting the constraints \eqref{vnu} and the Hamiltonian as
\be
C_1=\frac{e^{-2U}}{N} \left[ v ( v - \bar v ) - \frac{N^2}{r^2} \right]
\ ,\quad
C_2=dr - r v\ ,\quad
\ee
\be
H_{\rm WDW}  =
\frac{1}{N} C_2 ( C_2+ 2 r v) + e^{2U} C_1\ ,
\ee
and using \eqref{poisbra}, it is straightforward to check that the algebra
of constraints is first class,
\bse
\be
\{C_1,u\}=-i\frac{k N}{4 r^2} u\ ,\quad
\{C_1,E\}=-i\frac{k N}{4 r^2} E\ ,\quad
\{C_1,e\}=0
\ee
\be
\{C_2,u\}=\frac{N}{4 r}u\ ,\quad
\{C_2,E\}=\frac{N}{4 r} E\ ,\quad
\{C_2,e\} = \frac{N}{2 r} e
\ee
\be
\{C_1,C_2\} = -\frac{N}{2 r} C_1\ ,\quad
\{C_1, H_{\rm WDW}\} = \{C_2, H_{\rm WDW}\} = 0\ .\quad
\ee
\ese
Thus, the Hamiltonian system \eqref{wdw} can be further reduced
to the 8-dimensional symplectic quotient
\be
\label{tc1c2}
T^*( \IR^+ \times\mathcal{M}_3) \,//\,
\{ e = E = u = C_1 = C_2 = 0\}\ .
\ee
This is the habitat for the G\"odel and Eguchi-Hanson solutions
\eqref{godel} and \eqref{eguchi}. Note that this phase
space is also invariant under $\IR\times G_4 \times H_5$.

It is worth noting that the phase space \eqref{tc1c2}
can be further reduced with
respect to the second class constraints \eqref{c3c4}. In this
subspace, the Noether charge matrix is nilpotent of degree 7, and for
$k=-i$ the symmetry is enhanced to $SU(2)\times SU(2)$
(as the condition $C_4=0$ is equivalent to  $a/b=\pm i k$). This subspace
contains the G\"odel solution \eqref{godel}, as well as the non-spinning
BMPV solution \eqref{bmpv} with $j=\chi=0$.

\subsubsection*{The semi-classical radial wave function of G\"odel-type
black holes}
The five constraints
\be
e = E = u = C_1 = C_2 = 0
\ee
are solved in the sector
\be
Y_+ = q\ ,\quad E_k = k
\ee
by setting $p_a=\pa_{\phi^a}S_{q,k,\mathcal{S}}$, where
\be
\begin{split}
S_{q,k, \mathcal{S}} =& - i k\,e^{2U} + 2 i \frac{e^{-2U} r^2}{k} + \sqrt2 q \tau
+ k \left[ \sigma + \zeta^0 \tzeta_0 + \zeta^1 \tzeta_1 +
6 \tau \zeta_1^2 - 6 \tau^2 \zeta^0\zeta^1
+2 \tau^3 (\zeta^0)^2 \right]\\
&+  \mathcal{S}\left( \tzeta_0 + \tau \tzeta_1 +3 \tau^2 \zeta^1
- \tau^3\zeta^0, \tzeta_1+ 6 \tau \zeta^1
- 3 \tau^2 \zeta^0 \right)
\end{split}
\ee
and $ \mathcal{S}$ is an arbitrary function of two variables.
Thus, a basis of solutions of the quantum constraints
in the semi-classical approximation is given by
\be
\Psi_{q,k,p^0,p^1} \sim
\exp\left[ i S_{q,k,0} + i p^0
\left( \tzeta_0 + \tau \tzeta_1 +3 \tau^2 \zeta^1 - \tau^3\zeta^0\right)
+ i p^1 \left( \tzeta_1+ 6 \tau \zeta^1 - 3 \tau^2 \zeta^0\right)
\right]
\ee
Note that the wave function is a holomorphic function of $\tau$,
as required by the condition $e=0$, and that it flattens out in the
near horizon region where $a,b\to 0$.

\subsection{The Gutowski-Reall black hole}

\subsubsection*{Constraint analysis}
We now turn to the case of gauged supergravity, and study the consequences
of the natural generalization of the constraints \eqref{eEu}
to the gauged case,
\be
\label{eEut}
\utilde{e} = \utilde{E} =  \utilde{u} = 0 \ ,
\ee
In the presence of gauging, the constraints
\eqref{eEut} no longer commute with the Hamiltonian, but imply secondary
constraints. In particular,
\be
\label{hutilde}
\{H _{\rm WDW}, \utilde{u}\} = \frac{3 e^{-U}}{4 \ell \, \tau_2^{3/2}\, r}
\left( \tau_2 - i \tau_1\right) \,
\left( p_r + \frac{p_U}{r} - \frac{2ik e^{2U}}{r} \right)
\ee
\be
\label{hetilde}
\{H _{\rm WDW}, \utilde{e}\} = \frac{3 N e^{-2U}}{2 \ell^2 \tau_2^3}
\,\left( \tau_2 - i \tau_1\right) \,
\left( 3\sqrt{3} N(\tau_2- i \tau_1) -  e^U \tau_2^{3/2} \ell \utilde{\bar E}
\right)
\ee
Thus, we impose
\be
\label{eqt1}
C_0 \equiv \tau_1 + i \tau_2 = 0
\ee
which ensures the vanishing of both \eqref{hutilde} and \eqref{hetilde}.
This condition is in fact an integrated version of the
condition $\utilde{e}=0$ where the
integration constant has been fixed unambiguously.
Enforcing \eqref{eqt1}, the vanishing of
\be
\label{preq}
\{H _{\rm WDW},\utilde{E}\}
= \frac{\sqrt3 N e^{-U}}{4 \ell r \sqrt{\tau_2}}
\, \left(
p_r + \frac{p_U}{r} + \frac{2ik e^{2U}}{r} \right) = 0
\ee
implies the same constraint $C_2$ as in the ungauged case \eqref{vnu},
\be
\label{C2}
C_2 \equiv r' - r U' - i k \frac{N e^{2U}}{2 r}=0 \ .
\ee
Finally, requiring that $H _{\rm WDW}=0$ on the constraint locus requires that
\be
\label{pueq}
p_U = -2 i \,k \,e^{2U} - \frac{4i}{k} e^{-2U} r^2
+ \frac{6 \sqrt2\,e^{-2U} r^2}{\ell k } \left( p_{\tzeta_1}+ k \zeta_1 \right)
\ee
Expressing the derivative $\zeta^{1'}$ in terms of the charge $p^1=E_{p^1}$,
and $p_U$ in terms of $\utilde{v}$, this may be rewritten as
\be
\label{Ct1}
\utilde{C}_1 \equiv i k \utilde{v}-N\,e^{-2U}
- i\sqrt{\frac{3}{2}} \frac{N e^{-2U}}
{\ell} p^1 = 0\ .
\ee
In the limit $\ell \to \infty$, this reduces to the
condition $C_1=0$ in \eqref{vnu}. One may check that the
five conditions
\be
\label{fivec}
C_0 = \utilde{E} =  \utilde{u} = \utilde{C}_1 = C_2 = 0
\ee
commute, and therefore give a consistent first-class reduction of the phase
of $\IR\times U(1)\times SU(2)$ symmetric solutions of
gauged supergravity. The Hamilton-Jacobi functions satisfying
the constraints $\utilde{E} =  \utilde{u} = \utilde{C}_1 = C_2 = 0$
on the locus $C_0=0$ are
\be
\begin{split}
S_{k, \mathcal{S}} =&
-i k\, e^{2U}  + \frac{2i}{k} r^2\,e^{-2U} \left( 1 + \frac{3 i \sqrt2 k}{\ell}
 \zeta_1 \right)\\
& + k \left(  \sigma + \tzeta_0 \zeta^0 + \tzeta_1 \zeta^1\right)
+  \mathcal{S}\left( \tzeta_0, \tzeta_1 - \frac{3 \sqrt 2}{k\ell} r^2\,e^{-2U}  \right)\ ,
\end{split}
\ee
where $\mathcal{S}$ is an arbitrary function of two variables,
so that a basis of semi-classical wave functions are given by
\be
\Psi_{p^0,p^1}\sim
\exp\left[  i S_{k,0} + i p^0 \tzeta_0 
+ i p^1 \left(\tzeta_1 - \frac{3 \sqrt 2}{k\ell} r^2\,e^{-2U}  \right)\right]
\ee
Note that the semi-classical wave functions are now independent of $\tau$,
and flatten out near the horizon $a,b\to 0$.
As we now demonstrate shortly, this
reduced phase space does in fact contain the black hole
solution of \cite{Gutowski:2004ez}.

\subsubsection*{The Gutowski-Reall black hole}
Supersymmetric rotating black holes in minimal $D=5$ gauged supergravity
were constructed in \cite{Gutowski:2004ez}, and extended to general
$D=5$ gauged supergravity in \cite{Gutowski:2004yv}. In the
case of minimal supergravity, the solution is given by
\bse
\be
\cN=1\ ,\quad  a = \alpha \ell \sinh (\rho/\ell), \quad
b = 2 \alpha^2 \ell \sinh(\rho/\ell)\cosh (\rho/\ell),
\ee
\be
 f^{-1} = 1 + \frac{4 \alpha^2 -1}{12 \alpha^2 \sinh^2 (\rho/\ell)}\ ,
\quad
U_1= \frac{\sqrt{3}}{\ell} a^2
+ \frac{4\alpha^2-1}{2\sqrt3} \ell
\ee
\be
 \Psi = -2 \epsilon \alpha^2 \ell \sinh^2 (\rho/\ell) \left[ 1 + \frac{4
\alpha^2 -1}{4\alpha^2 \sinh^2 (\rho/\ell)} + \frac{(4\alpha^2 -1)^2}{96
\alpha^4 \sinh^4(\rho/\ell)} \right].
\ee
This may be supplemented with
\be
\tzeta_0 =0\ ,\quad \tzeta_1=\frac{3 i \sqrt{2}}{\ell} a^2\ ,\quad
\sigma=-\frac{i}{2} \alpha^2 \ell^2 \cosh(2\rho/l)
\left[ 1 - 2\alpha^2 + \alpha^2 \cosh(2\rho/l) \right]
\ee
\ese
so as to provide a solution of the motion on $\IR^+ \times G_{2(2)}/SO(4)$
with $E_{p^0}=E_{p^1}=0$ and $E_k=-i$. The Noether charges
for the symmetry group unbroken by the gauging are given by
\bse
\label{GRnoe}
\be
Y_+=-\frac{3 i  R_0^2}{4\sqrt{2}}\left(1 + \frac{R_0^2}{2\ell^2} \right)
\ ,\quad
F_{q_0}=
-\frac{3i R_0^4}{4\sqrt2 \ell} \left( 1+\frac{2R_0^2}{3\ell^2} \right)
\ee
\be
H-2Y_0 = \frac{3\ell^2}{16} \left( 1 + 2 \frac{R_0^2}{\ell^2}
- 3 \frac{R_0^4}{\ell^4} \right)\ ,\quad
E_{q_0}=E_{q_1}=0
\ee
\ese
where the parameter $\alpha$ was traded for
\be
R_0 = \ell \sqrt{ \frac{4\alpha^2-1}{3} }\ .
\ee
The charges $Y_+$ and $F_{q_0}$ are as usual
proportional to the charge and angular momentum,
\be
Q_e = \frac{\sqrt3 \pi R_0^2}{2G} \left( 1 + \frac{R_0^2}{2\ell^2} \right)
=\frac{2\pi i \sqrt2}{G \sqrt3}Y_+\ ,
\quad
J = \frac{3\epsilon \pi R_0^4}{8 G \ell} \left( 1+ \frac{2 R_0^2}{3\ell^2}
\right)=\frac{i\pi}{\sqrt2 G} F_{q_0}
\ee
Note however that the conserved charge $H-2Y_0$ differs from the ADM mass
\be
M = \frac{3\pi R_0^2}{4G} \left( 1 + \frac{3 R_0^2}{2\ell^2}
+ \frac{2 R_0^4}{3\ell^4} \right)
= \frac{\sqrt3}{2} |Q_e| +\frac{2}{\ell} |J|\ ,
\ee
although this could be easily rectified by adding a constant term in $\sigma$.
Finally, its Bekenstein-Hawking entropy is
\be
S_{BH} = \frac{\pi^2}{2G} \lim_{\rho\to 0}\,
\frac{a^2 \sqrt{b^2-f^3 \Psi^2}}{f^{3/2}}
= \frac{\pi^2 R_0^3}{2G}\,\sqrt{ 1 + \frac{3 R_0^2}{4\ell^2}}
\ee

We now check that this solution satisfies
the constraints \eqref{C2} and \eqref{Ct1}.
In terms of $\rho$-derivatives, the latter may be
rewritten as
\bse
\bea
U'/N &=& - \frac{i}{k} e^{-2U} -  \frac{i k e^{2U} }{2r^2}
+ \frac{3\sqrt2}{\ell} e^{-2U} \zeta^1\\
(r' - r U')/N &=& i k \frac{e^{2U}}{2 r}
\eea
\ese
Translating to the variables $a,b,\cN,\Psi,U^1$ using \eqref{abtoUr},
the conditions $\utilde{E}=0$ and $\utilde{u}=0$
become, respectively,
\bse
 \bea
\label{eqEb}
a^2 U^{1'} + i k b \cN  U^1 &=& \frac{2\sqrt{3}}{\ell} f^{-1} a^2 b \cN  \\
\label{equ}
a^2 U^{1'} - i k b \cN  U^1 &=& -\frac{f}{\sqrt3} \left( a^2 \Psi'
- \cN  b \Psi \right)
\eea
while the conditions \eqref{Ct1} and \eqref{C2} become
\be
\label{eqbU}
\frac{b'}{\cN} + i k \frac{b^2}{2a^2} - \frac{1}{ik}
= \frac{2\sqrt{3}}{\ell} U^1\ ,\qquad
i k b \cN =2 a a'
\ee
\ese
These four equations agree with Eq. (3.1),(3.15),(3.16),(3.17)
of \cite{Gutowski:2004yv} when $k=-i$ and in the gauge $\cN=1$,
while the constraint \eqref{eqt1}
is identical with the ansatz Eq. (3.11) in this same reference.
Moreover, solving \eqref{eqbU} for $U^1$ and plugging back in
\eqref{eqEb}, we find Eq. (3.12) in \cite{Gutowski:2004yv}
\be
f^{-1} = {\ell^2 \over 12 a^2 a'}
\left(4 (a')^3+7aa'a''-a'+a^2 a''' \right) \ .
\ee
Thus, we have recovered the Gutowski-Reall solution via algebraic 
considerations in the one-dimensional reduced model. 
We note that the ``flat limit'' $\ell\to\infty, \alpha\to\frac12$
keeping $R_0$ and $R$ fixed leads back to the
the BMPV black hole \eqref{bmpv} with zero angular momentum.
It would be very interesting to find a more general BPS solution
where the electric charge $Q$ and the angular momenta $J_1,J_2$
can be varied independently. Most likely, this requires going beyond
the cohomogeneity one case studied in this paper.

\subsection{Other solutions of gauged supergravity}
We first note that the $AdS_5$ vacuum solution is obtained by
taking $\alpha=\frac12$, and has $\utilde{\bar e}=0$ 
in addition to the five constraints in \eqref{fivec}.

Secondly, the near horizon
geometry is obtained by taking the limit $\rho\to 0$. Dropping
an irrelevant additive constant  in $\sigma$, we find
\be
\cN=1\ ,\quad a=\alpha \rho\ ,\quad b=2\alpha^2 \rho, \ ,\quad
f = \tau_2 = i \tau_1 = \frac{12 \alpha^2}{(4\alpha^2-1)\ell}\rho^2\ ,\quad
U^1 = \frac{(4\alpha^2-1)\ell}{2\sqrt3}\ ,\quad \nn
\ee
\be
\Psi=\frac{(4\alpha^2-1)^2\ell^3}{48 \alpha^2 \rho^2}\ ,\quad
\tzeta_0=0\ ,\quad \tzeta_1=\frac{3i\sqrt2 \alpha^2}{\ell} \rho^2\ ,\quad
\sigma = - i \alpha^2 \rho^2 
\ee
It has the same Noether charges as the full solution, and in addition
has $\utilde{\bar E}=0$. It would be interesting to connect this observation
to the enhancement of supersymmetry at the horizon \cite{Sinha:2006sh}.

Thirdly, we note that, upon relaxing the BPS condition,
a two-parameter family of $AdS_2 \times S^3$ geometries with
$E_{p^I}=0$ and $E_k=-i$ is allowed~\cite{Morales:2006gm}:
\be
\cN=1\ ,\quad a=\alpha \rho\ ,\quad b=2\alpha^2 \rho, \ ,\quad
f = \tau_2 = i \tau_1 = \frac{12 \alpha^2\beta}
{(4\alpha^2-1)\ell}\rho^2\ ,\quad
U^1 = \frac{(4\alpha^2-1)\ell}{2\sqrt3 \beta}\ ,\quad \nn
\ee
\be
\Psi=\frac{(4\alpha^2-1)^2\ell^3}{48 \alpha^2\,\beta^2\,\rho^2}\ ,\quad
\tzeta_0=0\ ,\quad \tzeta_1=\frac{3i\sqrt2 \alpha^2}{\ell} \rho^2\ ,\quad
\sigma = - i \frac{\alpha^2}{\beta} \left( 1 + 4 \alpha^2 (\beta-1) \rho^2 
\right)
\ee
This solution still satisfies $\utilde{E} =  \utilde{u}=C_0 = C_2=0$,
but has $\utilde{C}_1\neq 0$ unless $\beta=1$. Its electric charge and
angular momentum are now independent parameters,
\be
Y_+(\beta)= \frac{Y_+(1)}{\beta^2} - (\beta-1)\,\frac{3 i R_0^2}{4\sqrt2}
\left(1+\frac{3R_0^2}{\ell^2}\right)\ ,\quad 
F_{q_0}(\beta) = \frac{F_{q_0}(1)}{\beta^2} 
- (\beta-1)\,\frac{3 i R_0^4}{4\sqrt2 \ell} 
\left(1+\frac{3R_0^2}{\ell^2}\right)\ ,\quad 
\ee
where $Y_+(1)$ and $F_{q_0}(1)$ denote the values for the 
Gutowski-Reall black hole in \eqref{GRnoe}.
The Bekenstein-Hawking entropy is
\be
S_{BH} = \frac{\pi^2 R_0^3}{2G}\,\sqrt{ 1 + \frac{3 R_0^2}{4\ell^2}
+ (\beta-1) \left( 1 + \frac{3 R_0^2}{\ell^2} \right)}
\ee
It would be interesting to know if there exists a smooth interpolating
solution between this non-BPS extremal geometry and $AdS_5$ at infinity.

Finally, we note that BPS solutions of $\cN=1$ supergravity 
with naked singularities and closed time-like curves were 
constructed in \cite{Behrndt:1998ns,Klemm:2000vn}. In the non-rotating
case, their solution is given by 
\bse
\be
\cN=\left[ 1+ \frac{\rho^2}{\ell^2} 
\left( 1+\frac{q}{r^2}\right)^3\right]^{-1/4}\ ,\quad
a=b=\frac{\rho}{2\cN}\ ,\quad 
f=\cN^{-2} \left( 1+\frac{q}{r^2}\right)^{-1}\ ,\quad
\ee
\be
\tau_1= - i \left( 1+\frac{q}{r^2}\right)^{-1}\ ,\quad
\Psi=U^1=\tzeta_0=\tzeta_1=0\ ,\quad \sigma=-\frac{i}{4} \rho^2
\ee
\ese
Its conserved charges are
\be
Y_+=-\frac{3i}{4\sqrt2} q\ ,\quad E_k=-i\ ,\quad 
E_{p^I}=E_{q_I}=F_{q_0}=H-2Y_0=0
\ee
While the shifted quaternionic viel-bein does not seem to exhibit any
particular structure, the unshifted viel-bein satisfies
\be
u = \bar u = E = \bar E = 0\ ,\quad v=\bar v + N e^{-2U}\ .
\ee
Thus, for this type of solution the contributions $\cX^a B_\mu \gamma^\mu$
and $g \cX^a$ in \eqref{dchil} have to cancel. It is straightforward to see 
that the four conditions $u = \bar u = E = \bar E = 0$ commute with the 
Lagrangian on the constraint locus, however they are not first class 
since $v-\bar v$ does not vanish. It would be desirable to clarify the 
nature of this type of BPS solutions.

\section{Discussion}
In this work we took the first steps in extending the algebraic
methods which have been so useful for studying for 4D black holes,
to the case of $D=5, \cN=1$ supergravity, with and without gauging.
In particular, we have constructed the non-linear sigma model
arising in the reduction of stationary solutions with a $U(1)$ isometry
to $D=3$, and identified the appropriate gauging. We further studied the
reduction to $D=1$ appropriate to solutions with $U(1)\times SU(2)$
isometries, and studied the algebra of conserved charges and supersymmetry
constraints. These have been illustrated on a number of known 
solutions in gauged and ungauged gravity, including the BMPV black hole
and its generalizations, the G\"odel and Eguchi-Hanson black holes,
and the Gutowski-Reall solution. 

In the process, we have found evidence for a new supersymmetric completion 
of the bosonic sigma model, distinct
from the one relevant for 4D BPS black holes, and traced its origin 
to the non-trivial behavior of the Killing spinors along the fibers
of the reduction. We have also found that the supersymmetric solutions
of the ungauged theory exist in two branches, only one of which seems
to subsist at finite $\ell$. It would be very interesting to see if
more general supersymmetric solutions of gauged supergravity are allowed,
where an arbitrary linear combination of the first and second rows of the
(shifted) quaternionic viel-bein vanishes, or more generally whether
new (SUSY or non-SUSY) solutions may be reached by transformations in 
$G_{2(2)}$ which commute with the gauging.

The eventual goal of our construction is to provide a general framework
to describe 5D solutions of gauged supergravity, particularly in the BPS 
sector. While we have concentrated on BPS black hole solutions, with 
$\IR\times S^3$ conformal boundary, it would be useful to fit the 
AdS black strings, studied e.g. in \cite{Chamseddine:1999xk,Klemm:2000nj,
Mann:2006yi,Brihaye:2007vm,Bernamonti:2007bu}, in our formalism.
More ambitiously, it would be very desirable to extend our methods
to the co-homogeneity 2 case (relevant for stationary solutions with 
$U(1)\times U(1)$ isometries or co-homogeneity 3 case (for solutions
with a single isometry), which may allow us to construct new
multi-centered black hole or black ring solutions in $AdS$. 
We hope to return to some of these problems in future work.

\acknowledgments
We are grateful to P. Gao, C. Gowdigere, D. Reichmann, H. Samtleben, 
S. Vandoren and  A. Waldron for valuable discussions. 
B.P. and M. B. are grateful to their co-respective institutions for
hospitality during the course of this work.
The research of B.P. is supported in part by ANR(CNRS-USAR)
contract no.05-BLAN-0079-01.  The
research of M.B. is supported by the Israel-US binational science
foundation, by an Israeli science foundation center of excellence
1468/06, by a grant from the German Israel project cooperation, by
Minerva, by the EU network RTN-2004-512194 and by the GIF.

\appendix

\section{More general nilpotent solutions of degree 2\label{nil2app}}
The nilpotency condition $Q^2=0$ allows for more general solutions than
the Taub-NUT vacuum \eqref{tnsol} with  $j=\mu=0$. In view of the fact
that solutions with $Q^2=0$ are associated to the minimal co-adjoint
orbit of $G_3$, and the conjectural relation to the generalized 
topological amplitude~\cite{Gunaydin:2006bz,piogao}, we briefly present 
them here, postponing their interpretation to future work

For a given value of
the moduli, such solution are uniquely determined by the charges $p^I,q_I,k$
subject to two conditions, e.g. at the identity of $G_{2(2)}/SO(4)$
\be
3p^0q_1-\sqrt{3} q_1^2+\sqrt{3} (p^1)^2+3 p^1 q_0= 0\ ,\quad
k^2 = - \frac{(p^1 q_0-p^0 q_1)(3 p^0 q_0 + p^1 q_1)^2}
{6\sqrt3 (p^0 p^1+q_0 q_1)}\ .
\ee
For example, setting $k=-i, p^0=\sqrt{2} \mu/\sqrt{1+\mu^2},
q_0=i \sqrt{2}\mu$, we find
\bse
\be
a=\frac{1}{\rho+\gamma}
\left(1-\frac{2(2+\mu^2)}{\sqrt{1+\mu^2}}\rho+4\rho^2\right)^{1/4}\ ,\quad
b=\left(1-\frac{2(2+\mu^2)}{\sqrt{1+\mu^2}}\rho+4\rho^2\right)^{-1/4}
\ee
\be
\cN = a^2 b \ ,\quad
f=\sqrt{\frac{1-4\rho^2+\mu^2(1-2\rho(\sqrt{1+\mu^2}+2\rho))}{1+\mu^2-4\rho^2}}
\ee
\be
\Psi=\frac{2\mu\rho}{1-2\rho\sqrt{1+\mu^2}}\ ,\quad
\tzeta_0 = \frac{\sqrt{2}\mu\rho}{2\rho-\sqrt{1+\mu^2}}\ ,\quad
\tau_1=U^1=\tzeta_1=0
\ee
\be
\sigma=2 i \rho \frac{ 2 \mu^4 \rho^2 +
(4\rho^2-1)(1+\mu^2+(2+\mu^2\rho)\sqrt{1+\mu^2})}
{(1+\mu^2-4\rho^2)(1-4(1+\mu^2)\rho^2)}
\ee
\ese
where $\rho$ is equal to the geodesic affine parameter.
This solution has conserved charges
\bse
\be
H=-\frac{2(2+\mu^2)}{\sqrt{1+\mu^2}}\ ,\quad
Y_0=\frac{3\mu^2}{\sqrt{1+\mu^2}}\ ,\quad
F_{q^0}=2i\sqrt2 \mu\ ,
\ee
\be
E_{p^0}=-\frac{2\sqrt2 \mu}{\sqrt{1+\mu^2}}\ ,\quad
E_{q_0}=2i\sqrt2 \mu\ ,\quad E_{p^1}=E_{q_1}=Y_+=0\ .
\ee
\ese
This solution asymptotes to Taub-NUT space at $\rho\to -\gamma$,
and has an orbifold singularity at $\rho\to\infty$.
Note that it carries no electromagnetic flux in 5 dimensions.

\end{document}